\documentclass[10pt]{iopart}

\usepackage{iopams,amsfonts,amssymb,amsthm} 
\usepackage{latexsym}
\usepackage{rawfonts}
\usepackage{multirow}
\usepackage{rotating}
\usepackage{graphicx}
\usepackage{subfigure}
\usepackage{xcolor}
\usepackage{fancybox}

\input{prepictex}
\input{pictex}
\input{postpictex}

\definecolor{Maroon}{rgb}{0.70,0.0,0.0}
\definecolor{Maroon1}{rgb}{0.40,0.0,0.0}
\definecolor{Brown}{rgb}{0.7,0.3,0}
\definecolor{Navy}{rgb}{0.3,0.0,0.4}
\definecolor{Green}{cmyk}{1,0,1,0.2}
\definecolor{Red}{cmyk}{0,1,1,0}
\definecolor{DarkRed}{cmyk}{0,1,1,0.6}
\definecolor{DarkBlue}{cmyk}{1,1,0,0.5}
\definecolor{DarkGreen}{cmyk}{1,0,1,0.65}
\definecolor{OrangeRed}{cmyk}{0,1,0.87,0}
\definecolor{RedOrange}{cmyk}{0,0.77,0.87,0}
\definecolor{Orange}{cmyk}{0,0.61,0.87,0}
\definecolor{Offwhite}{cmyk}{.07,.15,.15,0}
\definecolor{Offwhite2}{cmyk}{.04,.02,.03,0}

\newcommand{\q}{\quad}

\newcommand{\Ref}[1]{(\ref{#1})}

\newcommand{\IntN}{{\mathbb{Z}}}

\def\L{\left(}
\def\R{\right)}
\def\LH{\left[}
\def\RH{\right]}
\def\LA{\left\langle}
\def\RA{\right\rangle}
\def\LV{\left|}
\def\RV{\right|}

\def\C#1{{\cal #1}}

\def\fns{\scriptsize}

\begin{document}
\color{black}

\title[Compressed Lattice Knots]{The Compressibility of Minimal Lattice Knots}

\author{E J Janse van Rensburg$\dagger$\footnote[3]{To whom 
correspondence should be addressed (\texttt{rensburg@yorku.ca)}}
and A Rechnitzer$\ddagger$}

\address{$\dagger$Department of Mathematics and Statistics, 
York University\\ Toronto, Ontario M3J~1P3, Canada\\
\texttt{rensburg@yorku.ca}}

\address{$\ddagger$Department of Mathematics, 
The University of British Columbia\\
Vancouver V6T~1Z2, British Columbia , Canada\\
\texttt{andrewr@math.ubc.ca}}

\begin{abstract}
The (isothermic) compressibility of lattice knots can be examined 
as a model of the effects of topology and geometry on the compressibility
of ring polymers.  In this paper, the compressibility of minimal
length lattice knots in the simple cubic, face centered 
cubic and body centered cubic lattices are determined.  
Our results show that the compressibility is generally
not monotonic, but in some cases increases with pressure.
Differences of the compressibility for different knot types show that
topology is a factor determining the compressibility of 
a lattice knot, and differences between the three lattices show that 
compressibility is also a function of geometry.
\end{abstract}


\maketitle

%%%%%%%%%%%%%%%%%%%%%%%%%%%%%%%%%%%%%%%%%%%%%%%%%%%%%%%%%%%%%%%%%%%%%%
%%%%%%%%%%%%%%%%%%%%%%%%%%%%%%%%%%%%%%%%%%%%%%%%%%%%%%%%%%%%%%%%%%%%%%
%%%%%%%%%%%%%%%%%%%%%%%%%%%%%%%%%%%%%%%%%%%%%%%%%%%%%%%%%%%%%%%%%%%%%%
%%%%%%%%%%%%%%%%%%%%%%%%%%%%%%%%%%%%%%%%%%%%%%%%%%%%%%%%%%%%%%%%%%%%%%
%%%%%%%%%%%%%%%%%%%%%%%%%%%%%%%%%%%%%%%%%%%%%%%%%%%%%%%%%%%%%%%%%%%%%%
\section{Introduction}

The compressibility of polymer colloids, melts, and also of biopolymers,
is an important physical property which affects rheology and other
physical properties \cite{NK59,ZK03}. The compressibility of linear
polymers is monotonic non-increasing with pressure (see for
example figure I in reference \cite{NK59} or figures 15 and 16 in
reference \cite{ZK03}), and typically approaches zero with increasing
pressure. 

In this paper, we will examine a model of the compressibility of 
tightly knotted ring polymers, where entanglements becomes an
important factor in addition to geometry in determining the 
thermodynamic properties of the polymer.  We do so by examining
the compressibility of a lattice model of a knotted ring polymer.

Lattice polygons were introduced as a model of polymer entropy in ring
polymers \cite{H61,deG79}.  In three dimensions a ring polymer
may be knotted, and it is known that the topological properties
of ring polymers have an important effect on entropy \cite{FW61,D62}.
Lattice knots are now a standard model for the polymer entropy problem 
in knotted ring polymers \cite{P89,SuW88,JvR99} and these objects
have been the subject of numerous studies over the last two decades 
\cite{MW86,JvRP95,V95,OJ98,JvR08,MMRS09,DIASV09,PDSAV10}.  One
of the advantages in models of lattice knots is that they are 
effective numerical models of simulating the effects of 
topological constraints (knotting) on the properties of ring
polymers, and although one may not obtain direct quantitative results 
for real ring polymers, qualitative results may be examined to 
gain insight into the physical properties of ring polymers generally.

The entropy of a ring polymer depends on its length (number of
monomers) and on knots along the backbone of the polymer.  Ring polymers
which are too short cannot accommodate a knot, and one expect for
each knot type there to be a minimal length which will allow the
knot to be realised in the polymer.  This situation is modeled by
minimal length lattice knots \cite{D93,D94}, which are idealised models
of knotted ring polymers of minimal length (or equivalently, 
maximal thickness).

\begin{figure}[h!]
\centering
\parbox{1.2in}{%
 \includegraphics[scale = 0.1] {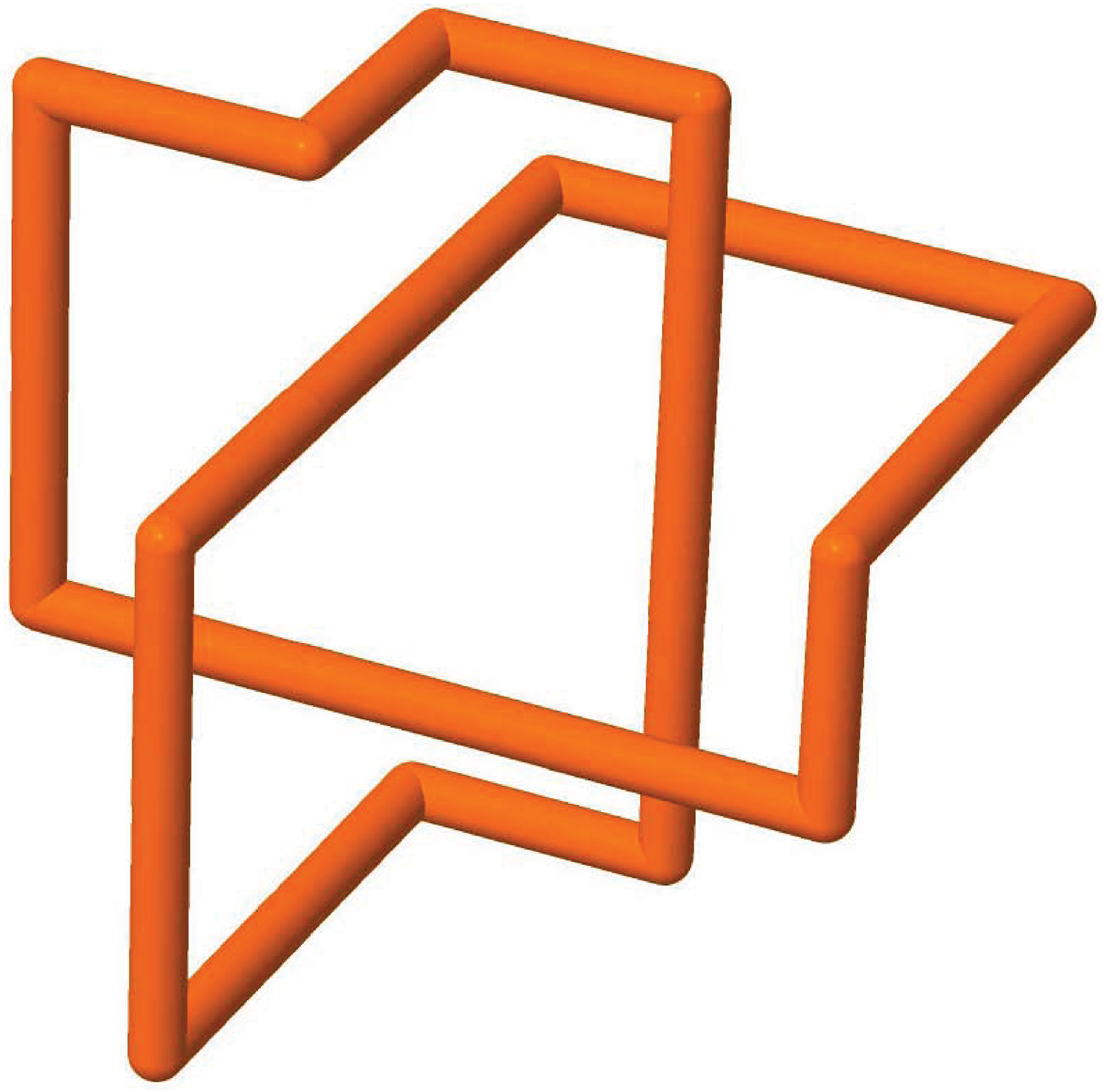}%
   %\caption{One}%
   \label{fig:subfig1}}%
\hspace{11mm}%
\begin{minipage}{1.2in}%
  \includegraphics[scale = 0.11] {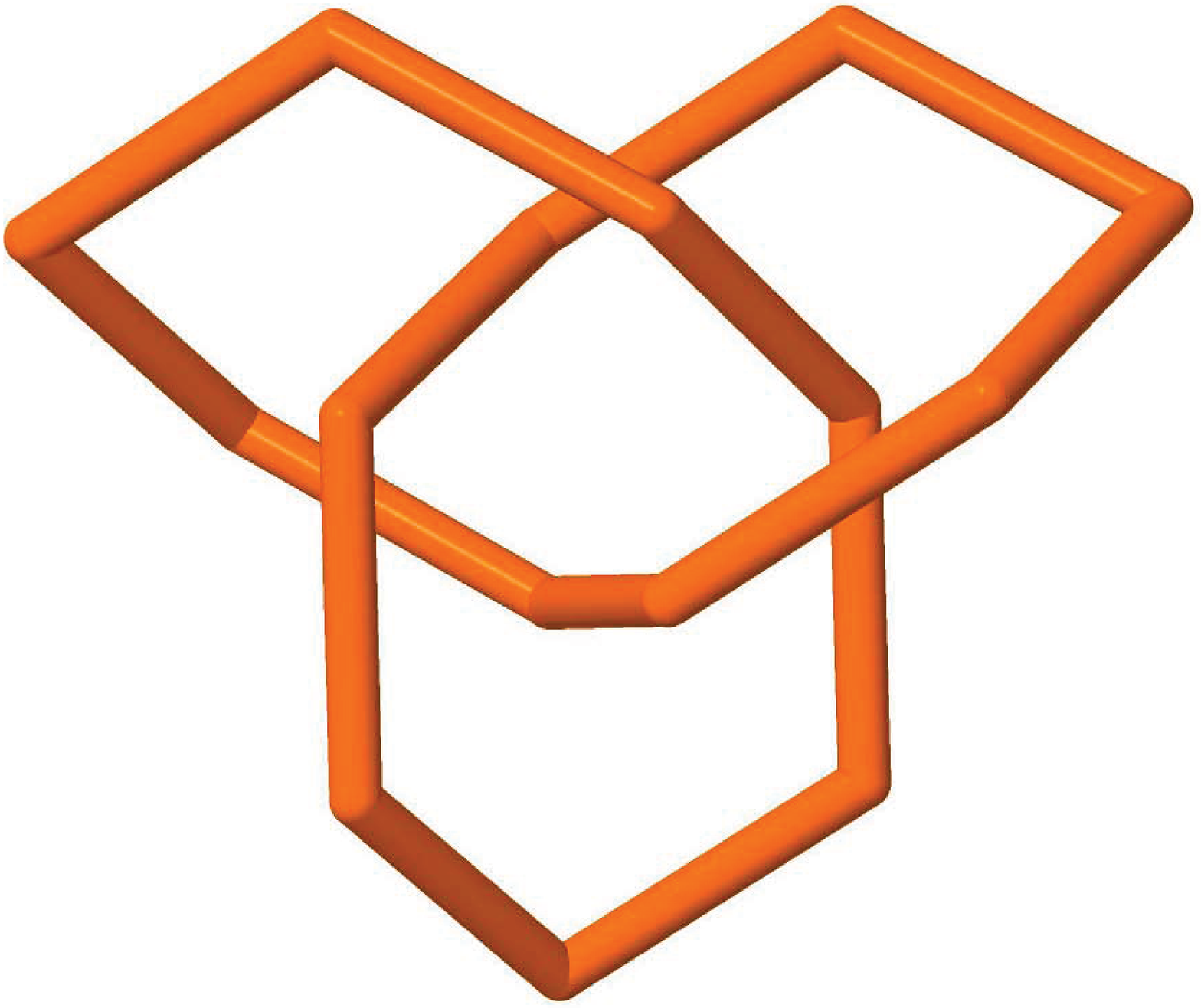}%
    %\caption{Second}%
    \label{fig:subfig2}%
\end{minipage}%
\hspace{12mm}%
\begin{minipage}{1.2in}%
  \includegraphics[scale = 0.11] {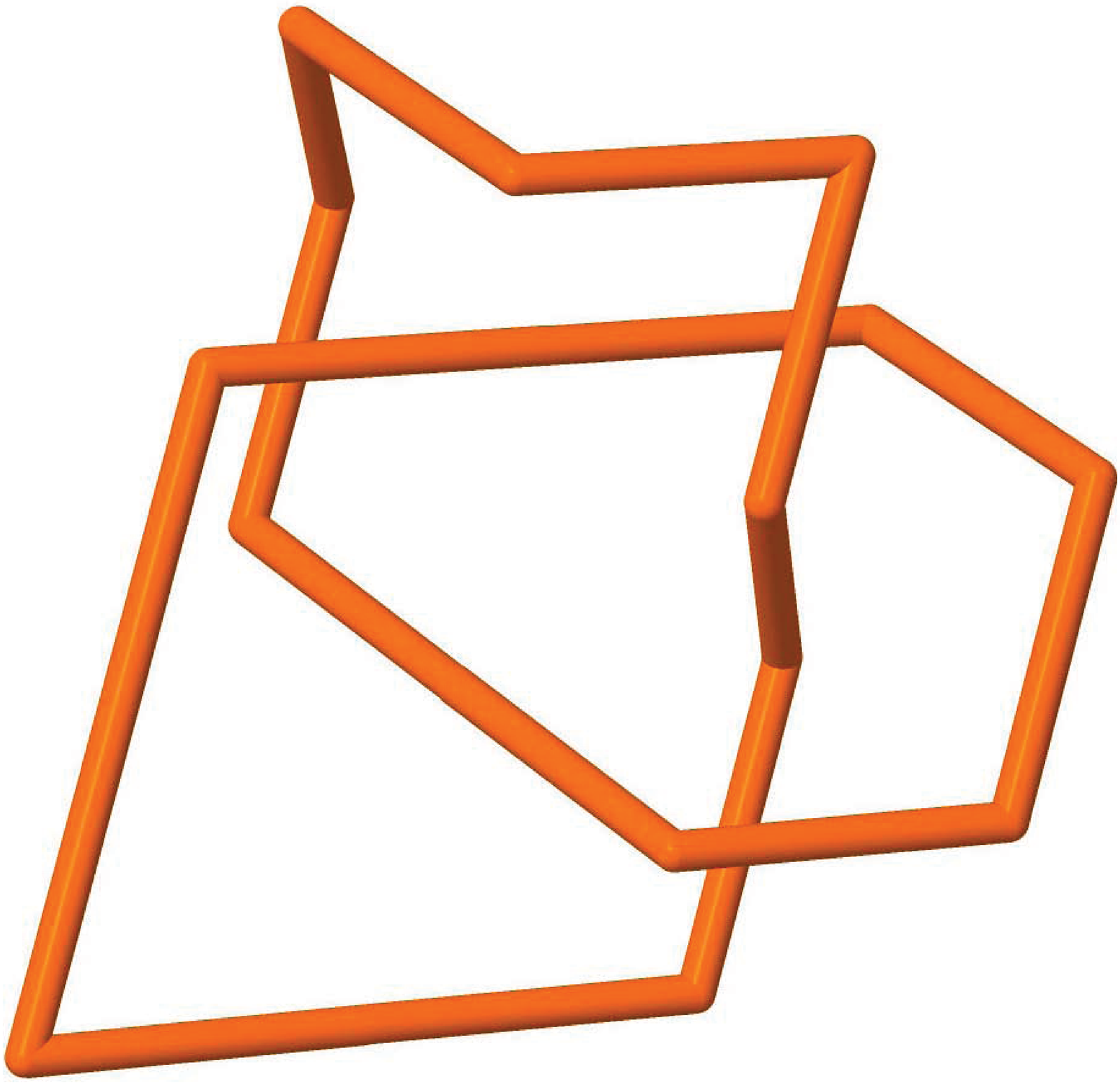}%
    %\caption{Second}%
    \label{fig:subfig3}%
\end{minipage}%
\caption{Examples of minimal length lattice trefoils 
(knot type $3_1^+$ in the standard knot tables \cite{BZ85}) 
in three dimensional cubic lattices.  These are embeddings 
in the simple cubic lattice (left), the face centered cubic 
lattice (middle), and the body centered cubic lattice (right).}
\label{FIG1} %%ZXZ[FIG1]
\end{figure}

Examples of minimal length lattice knot in the simple cubic,
face centered cubic, and body centered cubic lattices are illustrated 
in figure \ref{FIG1}. 
%%%%%%%%%%%%%%%%%%%%%%%%%%%%%%%%%%%%%%%%%%%%%%%%
%%%%%%%%%%%%%%%%%%%%%%%%%%%%%%%%%%%%%%%%%%%%%%%%%%%%%%

Minimal length embeddings of lattice knots have residual conformational
entropy (in addition to entropic contributions from translational 
and rotational degrees of freedom).  For example, it is known that 
there are $75$ distinct symmetry classes of right handed lattice 
trefoil knots, which expand to $1664$ distinct polygons when one 
accounts for rotational degrees of freedom as well \cite{DIASV09}.

The residual conformational degrees of freedom of minimal length
lattice knots may seem a surprising feature of these models, and
are known to be very dependent on the particular lattice \cite{JvRR11}.
However, minimal length ring polymers will similarly have some
residual conformational degrees of freedom, even if tightly knotted.
This would be because the discrete values of angles between adjacent
bonds joining the monomers will admit different local 
geometric arrangements in the embedding of the polymer in 
free space.  These arrangements will contribute to the entropy 
of the polymer, which is generally a function of the local geometry 
of the bonding (bonding angles, bond lengths).  It will also
be a function of the knot type if the polymer is knotted.
The overall topology of the knotted polymer will constrain the 
molecule even locally to entangle in particular ways to complete the
knot, in particular when it has minimal length.  This will 
determine in some measure the entropy of the polymer.

In other words, while the entropic properties of minimal length 
lattice polygons will not produce quantitative results of real
minimal length knotted ring polymers due to essential differences
between the geometry of the lattice and polymers in three space, 
lattice knots are nevertheless a useful model to examine the 
qualitative nature of the interplay between topology and entropy 
in ring polymers.

In this paper we examine the (isothermal) compressibility of lattice
knots as a model of the effects of pressure and topology on tightly
knotted polymers.  The compressibility will be defined by assuming
the lattice knot to be in a bath of solvent molecules which are excluded
from a volume about the lattice knot.  The response of the lattice
knot to increasing pressure in the solvent will be determined, and
its compressibility thus determined.

Our numerical approach will be to determine the compressibility 
from exact numerical data on minimal length lattice knots obtained 
by using the GAS-algorithm for lattice knots \cite{JvRR09,JvRR10}.  
An excluded volume containing the lattice knot will be defined, 
and changes in its mean size to an externally applied pressure 
will be used to calculate the compressibility of the lattice knot 
as it is compressed in this volume.  

Our approach defines an internal space close to the polygon 
which excludes solvent molecules, and increasing the ambient 
pressure will then compress the containing volume and the lattice
knot inside it.  We shall define two different volumes for this 
purpose, the first will be the smallest rectangular box containing 
the polygon, and the second will be a more close fitting envelope
which will be defined using slices defined by the two dimensional 
convex hulls of intersections of lattice knots with lattice planes.
The second volume is not necessarily convex, and will in many cases be smaller than the convex hull of the polygon, which is much harder to
determine. We shall repeat these calculations the three cubic lattices, 
namely the simple cubic lattice (SC), the face centered cubic 
lattice (FCC) and the body centered cubic lattice (BCC).

\begin{figure}[t!]
\centering
\input{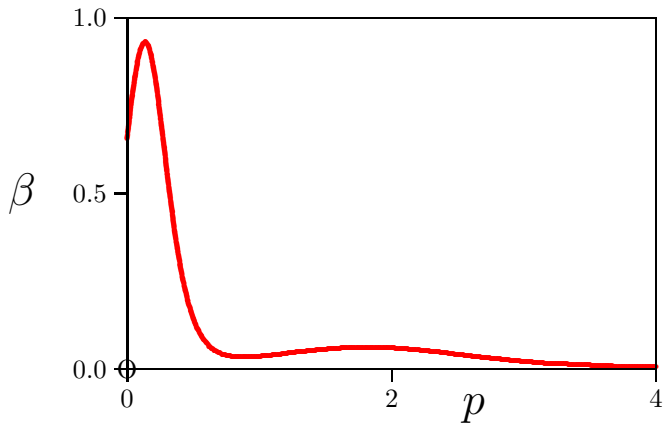}
\caption{Compressibility of the minimal lattice knot of
type $3_1^+$ in the SC lattice.}
\label{FIG2} %ZXZ[FIG2]
\end{figure}

In virtually all cases the lattice knots will be found to be
compressible in general.  Moreover, the compressibility will be found to
have complicated and unexpected dependence on the applied pressure in 
some knots types, even exhibiting increasing compressibility with
increasing pressure for some range of the applied pressure. For 
example, the compressibility $\beta$ of a lattice trefoil in 
a pressurised rectangular box in the simple cubic lattice is shown 
in figure \ref{FIG2}.
%%%%%%%%%%%%%%%%%%%%%%%%%%%%%%%%%%%%%%%%%%%%%%%%
%%%%%%%%%%%%%%%%%%%%%%%%%%%%%%%%%%%%%%%%%%%%%%%%%%%%%%
$\beta$ increases with small pressure to a peak, it then declines,
and reaches a local maximum at intermediate pressures before it
decreases to zero with large pressures.  This unusual shape for
the $(p,\beta)$ curve shows that lattice trefoils may become relatively
softer with increasing pressure in some pressure ranges -- this
kind of result would be dependent on both the lattice geometry and
the topology of polygon, and so would not necessarily give a qualitative
indication of the nature of real compressed ring polymers.  However,
the result suggests that this kind of behaviour is possible, as
it is observed to some extend in all three the lattices we
considered.  Generally, however, different curves will be obtained 
for $\beta$ in the other lattices.

We organised our paper as follows:  In section 2 our methods and
definitions are presented and explained.  In section 3 we present
results and discuss our findings.  We conclude the paper in section 4.

%%%%%%%%%%%%%%%%%%%%%%%%%%%%%%%%%%%%%%%%%%%%%%%%%%%%%%%%%%%%%%
%%%%%%%%%%%%%%%%%%%%%%%%%%%%%%%%%%%%%%%%%%%%%%%%%%%%%%%%%%%%%%
%%%%%%%%%%%%%%%%%%%%%%%%%%%%%%%%%%%%%%%%%%%%%%%%%%%%%%%%%%%%%%
\section{The Compressibility of Minimal Length Lattice Knots}

The SC lattice will be realised as $\IntN^3$ with basis
$\{ (1,0,0),(0,1,0),(0,0,1)\}$ and edges of length $1$.
The FCC lattice has basis $\{ (\pm 1,\pm 1, 0),(\pm 1, 0 , \pm 1),
(0,\pm 1,\pm 1)\}$ for all possible choices of the signs, and edges
between adjacent vertices of length $\sqrt{2}$, while the BCC will 
have basis $\{ (\pm 1, \pm 1, \pm 1) \}$ for all possible choices
of the signs, and edges of length $\sqrt{3}$.

Polygons in the SC, FCC and BCC lattices are sequences of vertices 
and edges $\{v_0,e_1,v_1,e_2,v_2,\ldots,v_{n-1},e_n,v_n\}$ where
$v_0 = v_n$ and all the vertices $v_i$ are distinct.  Then end-vertices
of edges $e_i$ are $v_{i-1}$ and $v_i$.  The \textit{length} if a polygon
$\omega$ is the number of edges it contains, and the \textit{chemical
length} will be the length times the length of edges:  For example,
if a polygon has length $n$, then its chemical length in the 
SC lattice is $n$, in the FCC lattice is $n \sqrt{2}$ and in the
BCC lattice is $n\sqrt{3}$ since edges have length $1$, $\sqrt{2}$ 
and $\sqrt{3}$ in the SC, FCC and BCC lattices respectively.  
In other words, the chemical length coincides with the geometric 
length of the polygons.

Polygons will be considered to be identical if one is a translate of
the other. Define $\C{P}_K$ to be the set of minimal length lattice 
polygons of knot type $K$ and length $n_K$.  The cardinality of
$\C{P}_K$ is the number of polygons of knot type $K$ and minimal
length, and this is denoted by $p_L(K) = \LV \C{P}_K \RV$ in the
lattice $L$.  For example, $p_{SC} (0_1) = 3$, $p_{FCC} (0_1) = 8$
while $p_{BCC} (0_1)=12$.  The minimal lengths of the unknot in the
SC lattice is $4$, while in the FCC  lattice it is $3$ and in 
the BCC lattice it is $4$.  

In order to determine the compressibility of lattice knots, it
is necessary to define the volume they occupy.   An upper bound on 
the excluded volume of a lattice knot is the minimal volume 
rectangular box with sides parallel to the $X$-, $Y$- and 
$Z$-axes containing it.  A lower bound can be obtained by noting
that a particular lattice knot $\omega$ should occupy at least the sites
along its length, and also the intersection of the minimal rectangular
box containing it with the union of the Wigner-Seitz cells along its 
length. This lower bound is perhaps too small, as interstitial space
surrounded by vertices of the lattice knot should also be part
of the excluded volume.  Hence, we choose to require the excluded
volume of a lattice knot to be a convex region containing the knot.

In the first instance, the smallest rectangular box in the lattice
containing the lattice knot was taken as an excluded volume $V_b$
about the knot.  A second, smaller excluded volume $V_e$ is defined
by slicing the lattice knot with lattice planes, computing convex 
hulls in these planes, and then gluing them in slabs together
to find a more tightly fitting volume about the lattice knot. We
call this volume the \textit{the averaged excluded volume} $V_e$ of the 
lattice knot.  A more careful definition is given below.
In both cases we assume that solvent molecules pressurising the
lattice knot is excluded from the excluded volume $V_b$ and
$V_e$, and increases in pressure in the solvent will pressurise
the knot by exerting a pressure on the convex boundary of the
volume about the lattice polygon.

Once the volumes containing lattice knots have been defined, 
they can be pressurized in these volumes and their compressibility
determined.  In this study we focus on the compressibility
of lattice knots by using the minimal rectangular box
and averaged excluded volume. 

\subsection{The rectangular box volume of lattice polygons}

The volume of the smallest rectangular box with sides normal to
the lattice axes containing a lattice knot is its \textit{box volume} 
$V_b$.  This volume is a large overestimate of the volume occupied by 
the polygon, since it is likely to include large spaces near the 
eight corners of the box.  Observe that if $V_C$ is the volume of the 
convex hull of a polygon, then in general $V_C \leq V_b$, although 
both volumes are minimal convex shapes containing the polygon.

Define $p_L (v;K)$ to be the number of minimal length lattice polygons
in a lattice $L$, of knot type $K$, minimal box volume $v$.  Then
$\sum_v p_L (v;k) = p_L (K)$ is the total number of minimal length 
lattice polygons of knot type $K$.

One may check that $p_{SC} (0;0_1)=3$ but $p_{FCC}(1;0_1)=8$.
However $p_{BCC}(8;0_1)=6$ and $p_{BCC}(4;0_1)=6$, since there are two
classes of minimal length unknotted polygons in the BCC lattice, one
class of 6 elements in a minimal rectangular box of dimensions
$2\times 2\times 2$, and a second class of 6 elements in
a minimal rectangular box of dimensions $2\times 2\times 1$.
%%%%%%%%%%%%%%%%%%%%%%%%%%%%%%%%%%%%%%%%%%%%%%%%
\begin{figure}[t!]
\centering
\input{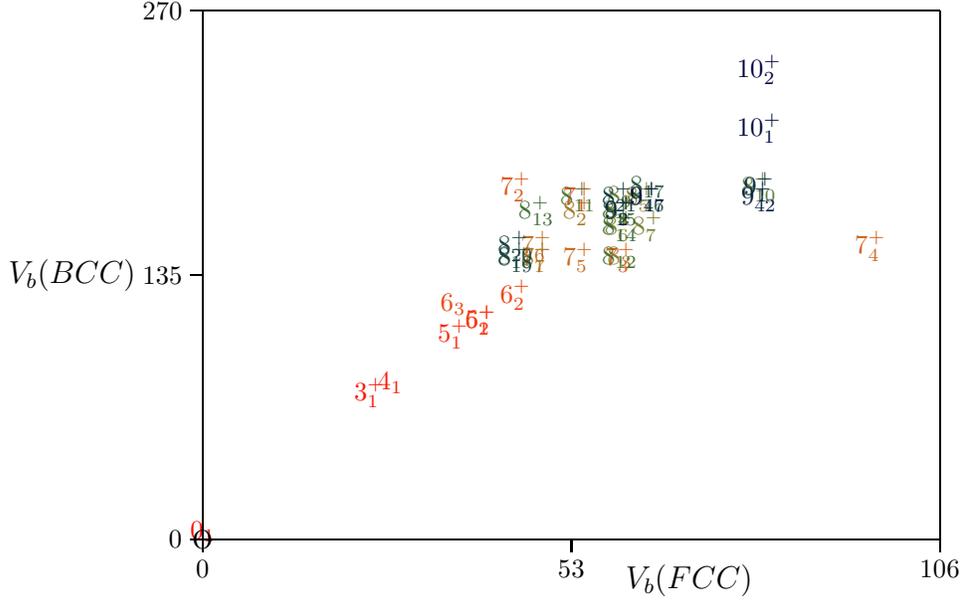}
\caption{A scatter plot of the minimum rectangular box volumes $V_b$
of knot types in the $FCC$ and $BCC$. The points are contained in 
a narrow wedge, and the volumes are strongly correlated.}
\label{GraphVbFCCBCC} %%ZXZ[GraphVbFCCBCC]
\end{figure}
%%%%%%%%%%%%%%%%%%%%%%%%%%%%%%%%%%%%%%%%%%%%%%%%%%%%%%

\subsection{The averaged excluded volume of lattice polygons}

Let $\omega$ be a lattice knot and $P_{z}$ be a lattice plane
normal to the $Z$-axis.  For some choices of $P_{z}$ the 
intersection $\omega \cap P_{z} = Q_\omega$ is not empty but contains
a subset of vertices from $\omega$.  Let $C_z$ be the 
convex hull of the points in $Q_\omega$, then $C_z$ is a 
(geometric) two dimensional polygon in the plane $P_z$.

Suppose that the (integer) values of $z$ giving a non-empty 
intersection $Q_\omega$ are $\{z_0,z_1,\ldots,z_m\}$ with
$z_{i-1} + 1 = z_i$ for $i=1,2,\ldots,m$.

Define the slab $S_z = C_z \times [z-1/2,z+1/2]$ for each $z=z_i$
with $i=1,2,\ldots,m-1$. In addition, define 
$S_{z_0} = C_{z_0} \times [z_0,z_0+1/2]$ and 
$S_{z_m} = C_{z_m} \times [z_m-1/2,z_m]$.

Define the volume $V_z$ about $\omega$ by taking the union of
these slabs:
\[ V_z = \bigcup_{i=0}^m S_{z_i} . \]
Observe that the vertices of $\omega$ are contained in $V_z$, but
that some edge may in fact be partially exposed.  Since excluded
volume effects in lattice polygons are defined in terms of
vertices avoiding one another, we still consider $V_z$
to constitute a volume containing $\omega$, but it fits
very tightly about the polygon in general.
%%%%%%%%%%%%%%%%%%%%%%%%%%%%%%%%%%%%%%%%%%%%%%%%
\begin{figure}[t!]
\centering
\input{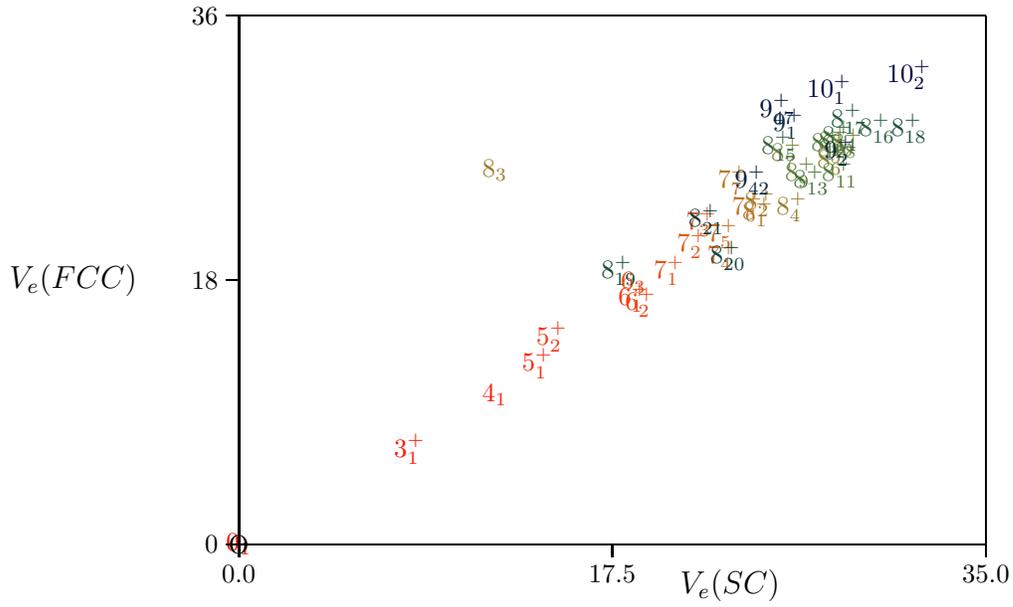}
\caption{A scatter plot of the minimum average excluded volumes $V_e$
of knot types in the $SC$ and $FCC$. The points, with the 
exception of $8_3$, are contained in a narrow wedge, and the volumes 
are strongly correlated.}
\label{GraphVeSCFCC} %%ZXZ[GraphVeSCFCC]
\end{figure}
%%%%%%%%%%%%%%%%%%%%%%%%%%%%%%%%%%%%%%%%%%%%%%%%%%%%%%
%%%%%%%%%%%%%%%%%%%%%%%%%%%%%%%%%%%%%%%%%%%%%%%%
\begin{figure}[h!]
\centering
\input{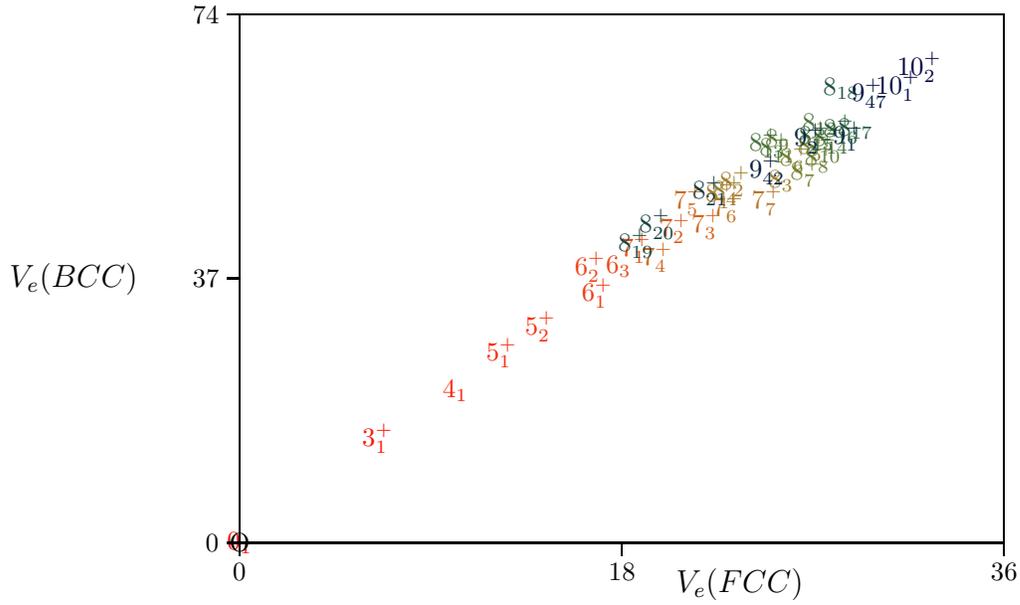}
\caption{A scatter plot of the minimum average excluded volumes $V_e$
of knot types in the $FCC$ and $BCC$. The points are contained in a 
narrow wedge, and the volumes are strongly correlated.}
\label{GraphVeFCCBCC} %%ZXZ[GraphVeFCCBCC]
\end{figure}
%%%%%%%%%%%%%%%%%%%%%%%%%%%%%%%%%%%%%%%%%%%%%%%%%%%%%%

Volumes $V_x$ and $V_y$ can be defined in this way by cutting 
$\omega$ by lattice planes normal to the $X$-axis or $Y$-axis
instead.  Taking the average gives the \textit{average excluded
volume} $V_e$ of $\omega$:
\[ V_e = (V_x + V_y + V_z)/3 . \]
Observed that $V_e \leq V_b$. One may also show that $V_e = 0$ 
for minimal length unknotted polygons in the SC, FCC and BCC lattices.

The values of $V_e$ in the three lattices are strongly correlated 
across knot types.  For example, in figure \ref{GraphVeSCFCC} a
scatter plot of $V_e$ in the SC and FCC is illustrated.
The minimal values of $V_e$ in the $FCC$ and $BCC$ are similarly
plotted in figure \ref{GraphVeFCCBCC} showing a strong correlation
between the these minimal volumes in the FCC and BCC lattices.

\begin{figure}[h!]
\centering
\input{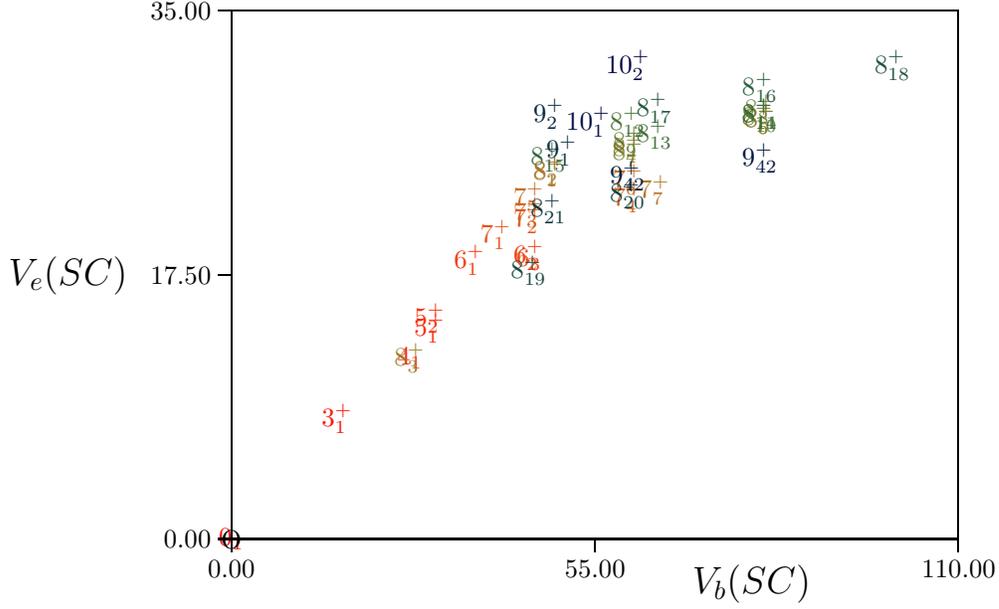}
\caption{A scatter plot of $\min V_e (K)$ and $\min V_b (K)$ in the SC
lattice for knot types $K$ as indicated.  The points are contained in 
a narrow wedge, and the volumes are strongly correlated.}
\label{VminSCSC} %ZXZ[VminSCSC]
\end{figure}

The minimal values of $V_b$ and $V_e$ are strongly correlated for
different knot types.  For example, if the minimum value of $V_b$ is
relatively small for a knot type $K$, then so is the minimum value
of $V_e$.  In figure \ref{VminSCSC} the minimal values of $(V_b,V_e)$
are illustrated in a scatter plot for the SC lattice.
%%%%%%%%%%%%%%%%%%%%%%%%%%%%%%%%%%%%%%%%%%%%%%%%
%%%%%%%%%%%%%%%%%%%%%%%%%%%%%%%%%%%%%%%%%%%%%%%%%%%%%%

\subsection{The compressibility of minimal length lattice polygons}

Consider a minimal lattice knot $\omega$ in a volume $v_\omega$
(which may be either the minimal box volume $V_b$, or the average 
excluded volume $V_e$).  If the lattice knot is pressurized by an 
external pressure $p$, then its statistical mechanics properties will
be described by its partition function, given by
\[ \C{Z}_L (K) = \sum_v p_L (v; K) e^{-p\,v} \]
where pressure and volume are in lattice units, and the Boltzmann
factor is put equal to unity.

For example, the partition functions of minimal unknotted polygons
with volumes $V_b$ are given by
\[ \C{Z}_{SC}(0_1) = 3,\q
   \C{Z}_{FCC}(0_1) = 8e^{-p},\q
   \C{Z}_{BCC}(0_1) = 6e^{-8p}+6e^{-4p} . \] 
If the volume $V_e$ is used instead, then
\[ \C{Z}_{SC}(0_1) = 3,\q
   \C{Z}_{FCC}(0_1) = 8,\q
   \C{Z}_{BCC}(0_1) = 12 . \] 
The free energy of these models van be computed in the usual
way: $\C{F}_L (K) = -\log \C{Z}_L (K)$.  For example, the
free energy of minimal length unknotted polygons in the
BCC lattice in the rectangular box volume $V_b$ is given by
\begin{equation}
\C{F}_{BCC} (0_1) = -\log \L 6e^{-8p}+6e^{-4p} \R. 
\label{eqnFBCC01} %ZXZ[eqnFBCC01]
\end{equation}
This is explicitly dependent on $p$, and shows that pressurising
minimal length unknotted BCC lattice polygons will lead to a response
by adjusting the average equilibrium volume occupied by the polygon.

In the other cases the polygon will not respond to increasing pressure, 
and it is incompressible.

The average (equilibrium) volume occupied by a minimal length 
lattice knot at pressure $p$ is obtained by
\[ \LA V \RA_L = \frac{\partial \C{F}_L(K)}{\partial p} . \]
The compressibility of the body is defined by
\[ \beta = - \frac{\partial \log \LA V\RA_L}{\partial p} , \]
and this definition shows that $\beta$ is the fractional change in 
expected volume due to an increment in pressure.  $\beta$ is
generally a function of temperature and pressure, and since
temperature is fixed in this model, this is in particular
the isothermal compressibility.

The average occupied rectangular box volume and compressibility of 
the minimal length unknot in the BCC lattice is
\[ \LA V_b \RA_{BCC} =  
\frac{4(1+2e^{-4p})}{1+e^{-4p}} . \]
When $p=0$ this gives $\LA V_b \RA_{BCC} = 6$ and if $p\to\infty$,
then $\LA V_b \RA_{BCC} = 4$.  The compressibility can be computed
directly:
\[ \beta_{BCC} (0_1) = \frac{4e^{-4p}}{(1+e^{-4p})(1+2e^{-4p})} . \]
For the minimal length unknot in the other lattices, and also for
the choice of $V_e$ in all cases, the compressibility of the unknot is zero
(that is, it is not compressible).
%%%%%%%%%%%%%%%%%%%%%%%%%%%%%%%%%%%%%%%%%%%%%%%%
\begin{figure}[h!]
\centering
\input{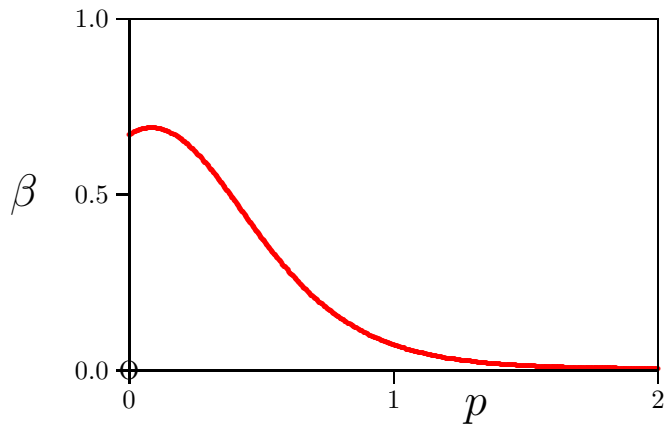}
\caption{Compressibility of the BCC minimal lattice knot of type $0_1$ in a rectangular
box of volume $V_b$.  Observe the global maximum at $p=0.08664\ldots$ where
$\beta=0.68629\ldots$.  If $p=0$, then $\beta=2/3$.}
\label{FIG2A} %ZXZ[FIG2A]
\end{figure}
%%%%%%%%%%%%%%%%%%%%%%%%%%%%%%%%%%%%%%%%%%%%%%%%%%%%%%
The compressibility of the unknot in the BCC lattice with volume
$V_b$ is plotted in figure \ref{FIG2A}.

The minimum rectangular box volume containing a lattice knot $K$ is 
its \textit{minimal rectangular box volume} (in the case $V_b$), or 
its \textit{minimal average excluded volume} (in the case of $V_e$).  
If a lattice knot is allowed to expand from its minimal
volume to its equilibrium volume at zero pressure, there is a change
in its free energy.  The maximum amount of useful work that can
be extracted from this expansion, if the process is reversible, 
is given by the difference in free energies between the states at 
$p=0$ (when the lattice knot is in an equilibrium state at zero pressure)
and the state were the lattice knot is confined to its minimal volume.

Since the free energy of the unknot is independent of the pressure
for the volume $V_e$, no useful work can be performed by a compressed
unknot in any of the lattices.  With the choice of volume $V_b$, the
one exception is in the BCC lattice where $\C{F}_{BCC} (0_1)$ is given 
by equation \Ref{eqnFBCC01}.  Direct calculation shows that the 
maximum amount of work that can be performed by letting this polygon 
expand reversibly from its compressed state is $\C{W}_{0_1} = 
\log 12 - \log 6 = \log 2 $.

%%%%%%%%%%%%%%%%%%%%%%%%%%%%%%%%%%%%%%%%%%%%%%%%%%%%%%%%%%%%%%%%%%%%%%
%%%%%%%%%%%%%%%%%%%%%%%%%%%%%%%%%%%%%%%%%%%%%%%%%%%%%%%%%%%%%%%%%%%%%%
\section{Compressibility of Lattice Knots: Numerical Results}

\subsection{GAS sampling of lattice knots}

In this study we have sampled minimal length lattice polygons
by implementing the GAS algorithm \cite{JvRR09,JvRR10}.  
The algorithm is implemented using a set of local elementary 
transitions (called ``atmospheric moves" \cite{JvRR08}) to 
sample along sequences of polygon conformations. The algorithm 
is a generalisation of the Rosenbluth algorithm \cite{RR55}, and 
is an approximate enumeration algorithm.

The GAS algorithm can be implemented in the SC lattice
on polygons of given knot type $K$ using the BFACF elementary
moves \cite{AC83,BF81} to implement the atmospheric moves
\cite{JvRR10,JvRR11}. This implementation is irreducible on classes
of polygons of fixed knot type \cite{JvRW91}.

Atmospheric moves in the FCC and BCC lattices were defined in
reference \cite{JvRR11}, and implemented using the GAS algorithm.
As in the SC lattice, these atmospheric moves are irreducible
on polygon classes of fixed knot type.

The implementation is described in detail in references
 \cite{JvRR10,JvRR11}).  Polygons of fixed knot type were sampled 
along sequences, tracking their length and metric properties.
Minimal length polygons were detected, classified by symmetry
class, stored by hashing and written to disk for later analysis.

The GAS algorithm was implemented efficiently using hash coding
which allowed elementary moves to be executed in $O(1)$ CPU time,
independent of length. This efficiency allowed us to perform
billions of iterations on knotted polygons in reasonable real 
time on desk top linux workstations. Simulations were performed
by sampling up to 500 GAS sequences, each of length $10^7$, to
search and collect minimal length polygons.  In most cases the
minimal length polygons were detected fairly quickly, although 
some knot types prove a little harder and required more CPU time.

Data on minimal knots were collated and analysed separately.  Minimal
volumes $V_b$ and $V_e$ were computed for symmetry classes of
the polygons and collected into expressions for the partition
functions, from which free energies and compressibility were computed.

\subsection{Compressibility of minimal length lattice knots in the SC lattice}

The partition function of minimal length lattice knots with volumes
given by $V_b$ and up to 7 crossings are given in 
table \ref{PFKnot-Box-SC}.
%%%%%%%%%%%%%%%%%%%%%%%%%%%%%%%%%%%%%%%%%%%%%%%%%%%%%%%%%%%%%%%%%%%%%%%%%%
\begin{table}[h!]
\begin{center}
 \begin{tabular}{||c||l||}
 \hline
  Knot & \multicolumn{1}{|c||}{$\C{Z}_{SC}(p)$} \\
  \hline  
$0_1$ &\fns $3$ \\
$3_1^+$ &\fns $1220{e^{-27p}}+432{e^{-18p}}+12{e^{-16p}}$\\
$4_1$ &\fns $2784{e^{-36p}}+864{e^{-27p}}$\\
$5_1^+$ &\fns $432{e^{-48p}}+624{e^{-36p}}+2160{e^{-32p}}+120{e^{-30p}}$\\
$5_2^+$ &\fns $40536{e^{-48p}}+4200{e^{-45p}}+10560{e^{-36p}}+2160{e^{-30p}}$\\
$6_1^+$ &\fns $2640{e^{-48p}}+216{e^{-45p}}+216{e^{-36p}}$\\
$6_2^+$ &\fns $7560{e^{-64p}}+5256{e^{-60p}}+3504{e^{-48p}}+96{e^{-45p}}$\\
$6_3$ &\fns $2208{e^{-64p}}+912{e^{-48p}}+432{e^{-45p}}$\\
$7_1^+$ &\fns $288{e^{-80p}}+252{e^{-64p}}+11808{e^{-60p}}+300{e^{-54p}}+468{e^{-48p}}+264{e^{-45p}}+3600{e^{-40p}}$\\
$7_2^+$ &\fns $22404{e^{-80p}}+2880{e^{-75p}}+13032{e^{-64p}}+85008{e^{-60p}}+11712{e^{-54p}}+22512{e^{-48p}}+10632{e^{-45p}}$\\
$7_3^+$ &\fns $240{e^{-45p}}$\\
$7_4^+$ &\fns $24{e^{-64p}}+60{e^{-60p}}$\\
$7_5^+$ &\fns $4392{e^{-64p}}+216{e^{-60p}}+96{e^{-48p}}+24{e^{-45p}}$\\
$7_6^+$ &\fns $10368{e^{-80p}}+4728{e^{-64p}}+1920{e^{-60p}}$\\
$7_7^+$ &\fns $168{e^{-80p}}+84{e^{-64p}}$\\
 \hline
 \end{tabular}
\end{center}
 \caption{Partition Functions: Minimal SC lattice knots
in a Pressurized Rectangular Box.}
  \label{PFKnot-Box-SC}  %ZXZ[PFKnot-Box-SC]
\end{table}
%%%%%%%%%%%%%%%%%%%%%%%%%%%%%%%%%%%%%%%%%%%%%%%%%%%%%%%%%%%%%%%%%%%%%%%%%%
%%%%%%%%%%%%%%%%%%%%%%%%%%%%%%%%%%%%%%%%%%%%%%%%%%%%%%%%%%%%%%%%%%%%%%%%%%
The partition functions of minimal lattice knots with volumes $V_e$
(average excluded volumes) are listed in table \ref{PFKnot-CVH-SC}
up to 5 crossing knots.  These expression are typically far
more complicated than those in table \ref{PFKnot-Box-SC}.
%%%%%%%%%%%%%%%%%%%%%%%%%%%%%%%%%%%%%%%%%%%%%%%%%%%%%%%%%%%%%%%%%%%%%%%%%%
%%%%%%%%%%%%%%%%%%%%%%%%%%%%%%%%%%%%%%%%%%%%%%%%%%%%%%%%%%%%%%%%%%%%%%%%%%
\begin{table}[h!]
\begin{center}
 \begin{tabular}{||c||l||}
 \hline
  Knot & \multicolumn{1}{|c||}{$\C{Z}_{SC}(p)$} \\
  \hline  
$0_1$ &\fns $3$ \\ \hline
$3_1^+$ &\fns $84{{e}^{-32p/3}}+24{{e}^{-65p/6}}+300{{e}^{-10p}}
+96{{e}^{-61p/6}}+240{{e}^{-59p/6}}+128{{e}^{-21p/2}}+48{{e}^{-28p/3}}$\\
$ $     &\fns $+168{{e}^{-31p/3}}+48{{e}^{-19/2p}}+72{{e}^{-29p/3}}+12{{e}^{-11p}}+72{{e}^{-103p/12}}+192{{e}^{-53p/6}}+48{{e}^{-26p/3}}$\\
$ $     &\fns $+24{{e}^{-35p/4}}+48{{e}^{-17/2p}}+24{{e}^{-25p/3}}
+24{{e}^{-101p/12}}+12{{e}^{-8p}}$\\ \hline
$4_1$   &\fns $48{{e}^{-155p/12}}+48{{e}^{-43p/3}}+192{{e}^{-79p/6}}
+96{{e}^{-151p/12}}+96{{e}^{-25p/2}}+96{{e}^{-175p/12}}+96{{e}^{-38p/3}}$\\
$ $     &\fns $+96{{e}^{-13p}}+192{{e}^{-173p/12}}+144{{e}^{-55p/4}}+192{{e}^{-77p/6}}+288{{e}^{-14p}}+48{{e}^{-59p/4}}+96{{e}^{-12p}}$\\
$ $     &\fns $+48{{e}^{-41p/3}}+144{{e}^{-40p/3}}+96{{e}^{-53p/4}}+96{{e}^{-167p/12}}+432{{e}^{-83p/6}}+192{{e}^{-85p/6}}$\\
$ $     &\fns $+48{{e}^{-157p/12}}+192{{e}^{-57p/4}}
+240{{e}^{-161p/12}}+192{{e}^{-27p/2}}+240{{e}^{-163p/12}}$\\ \hline
$5_1^+$ &\fns $96{{e}^{-29/2}}+48{{e}^{-173p/12}}+240{{e}^{-185p/12}}+264{{e}^{-91p/6}}+144{{e}^{-59p/4}}+72{{e}^{-187p/12}}$\\
$ $     &\fns $+192{{e}^{-61p/4}}+264{{e}^{-181p/12}}+120{{e}^{-175p/12}}+120{{e}^{-43p/3}}+96{{e}^{-44p/3}}+288{{e}^{-89p/6}}$\\
$ $     &\fns $+264{{e}^{-31p/2}}+204{{e}^{-15p}}+276{{e}^{-46p/3}}+144{{e}^{-179p/12}}+252{{e}^{-47p/3}}+168{{e}^{-95p/6}}+12{{e}^{-16p}}$\\
$ $     &\fns $+24{{e}^{-14p}}+48{{e}^{-169p/12}}$\\ \hline
$5_2^+$ &\fns $2496{{e}^{-109p/6}}+792{{e}^{-115p/6}}+72{{e}^{-47p/3}}+48{{e}^{-20p}}+48{{e}^{-179p/12}}+192{{e}^{-235p/12}}$\\
$ $     &\fns $+912{{e}^{-229p/12}}+1320{{e}^{-71p/4}}+1536{{e}^{-227p/12}}+24{{e}^{-181p/12}}+504{{e}^{33p/2}}+720{{e}^{-77p/4}}$\\
$ $     &\fns $+2232{{e}^{-35p/2}}+600{{e}^{-39p/2}}+1656{{e}^{-209p/12}}+1440{{e}^{-215p/12}}+72{{e}^{-61p/4}}+1584{{e}^{-113p/6}}$\\
$ $     &\fns $+768{{e}^{-67p/4}}+120{{e}^{-46p/3}}+696{{e}^{-205p/12}}+264{{e}^{-187p/12}}+2088{{e}^{-53p/3}}+384{{e}^{-59p/3}}$\\
$ $     &\fns $+1392{{e}^{-103p/6}}+192{{e}^{-119p/6}}+1152{{e}^{-17p}}+456{{e}^{-199p/12}}+408{{e}^{-49p/3}}+2184{{e}^{-56p/3}}$\\
$ $     &\fns $+72{{e}^{-15p}}+360{{e}^{-63p/4}}+72{{e}^{-185p/12}}+504{{e}^{-65p/4}}+1608{{e}^{-52p/3}}+1224{{e}^{-211p/12}}$\\
$ $     &\fns $+1728{{e}^{-221p/12}}+216{{e}^{-95p/6}}+3264{{e}^{-107p/6}}+2496{{e}^{-37p/2}}+384{{e}^{-197p/12}}+120{{e}^{-79p/4}}$\\
$ $     &\fns $+1728{{e}^{-223p/12}}+1056{{e}^{-101p/6}}+2916{{e}^{-18p}}+168{{e}^{-191p/12}}+312{{e}^{-97p/6}}+1200{{e}^{-75p/4}}$\\
$ $     &\fns $+1728{{e}^{-73p/4}}+24{{e}^{-59p/4}}+768{{e}^{-203p/12}}+336{{e}^{-193p/12}}+600{{e}^{-50p/3}}+2712{{e}^{-55p/3}}$\\
$ $     &\fns $+24{{e}^{-121p/6}}+1128{{e}^{-58p/3}}+24{{e}^{-44p/3}}+276{{e}^{-16p}}+1224{{e}^{-69p/4}}+96{{e}^{-91p/6}}$\\
$ $     &\fns $+120{{e}^{-31p/2}}+1200{{e}^{-217p/12}}+480{{e}^{-233p/12}}+936{{e}^{-19p}}$\\
  \hline
 \end{tabular}
\end{center}
 \caption{Partition Functions: Minimal SC lattice knots in pressurized in their
averaged excluded volumes.}
  \label{PFKnot-CVH-SC} %ZXZ[PFKnot-CVH-SC]
\end{table}
%%%%%%%%%%%%%%%%%%%%%%%%%%%%%%%%%%%%%%%%%%%%%%%%%%%%%%%%%%%%%%%%%%%%%%%%%%
%%%%%%%%%%%%%%%%%%%%%%%%%%%%%%%%%%%%%%%%%%%%%%%%%%%%%%%%%%%%%%%%%%%%%%%%%%
By examining the partition functions in tables \ref{PFKnot-Box-SC}
and \ref{PFKnot-CVH-SC} the compressibility and other properties
of the minimal length lattice knots can be determined exactly.

\subsubsection{Compressibility of $3_1^+$ in the SC lattice:}
The compressibility of the trefoil knot displayed in figure
\ref{FIG2} was obtained by choosing the volume $V_b$ in the 
SC lattice.  The partition function in this case is given by
\begin{equation}
\C{Z}_{SC} (3_1^+) = 1220{e^{-27p}}+432{e^{-18p}}+12{e^{-16p}}. 
\label{eqnZSC31} %ZXZ[eqnZSC31]
\end{equation}

The expected volume of the lattice knot at pressure $p$ is
\[ \LA V_b \RA
= \frac {3\L 2745e^{-27p}+648e^{-18p}+16e^{-16p}\R}{
305e^{-27p}+108e^{-18p}+3e^{-16\,p}}
\]
and the compressibility is explicitly given by
\[ \fl \beta = 
\frac {889380e^{-45p}+36905e^{-43p}+432e^{-34p}}{
 \L 2745e^{-27p}+648e^{-18p}+16e^{-16p} \R 
 \L 305e^{-27p}+108e^{-18p}+3e^{-16p} \R } . \]
Plotting this function for $p\in [0,4]$ gives the curve in figure
\ref{FIG2}.

The compressibility of $3_1^+$ is not monotone with increasing $p$ but 
first goes through a global maximum and then again later through
a local maximum at higher pressure.  Paradoxically, the lattice
knot becomes ``softer" when compressed for some ranges of the
pressure, in the sense that the fractional decrease in volume 
increases with $p$ in some instances.  This effect is likely due
to the choice of $V_b$ as an enclosing volume -- as the pressure
increase, the lattice knot starts to explore conformations which
expand into the corners of the containing box more frequently,
with the result that there is still space to expand into and 
compensate for the increasing pressure.  The maxima in figure \ref{FIG2} 
can be determined numerically.  The global maximum is located at
$p = 0.13557\ldots$ where $\beta = 0.92845\ldots$.  The local 
maximum is located at $p=1.82115\ldots$ where $\beta = 0.05888\ldots$.
The compressibility at zero pressure is $\beta = 0.65347\ldots$.

\begin{figure}[t!]
\centering
\input{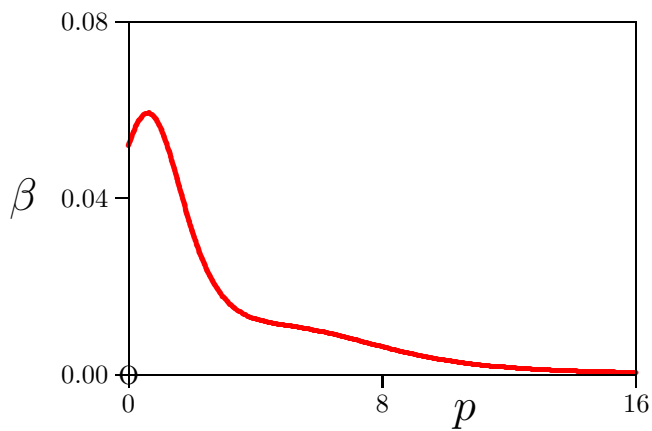}
\caption{Compressibility of the minimal lattice knot of
type $3_1^+$ with volume $V_e$ in the SC lattice.}
\label{FIG3} %ZXZ[FIG3]
\end{figure}

These results should be compared when the SC lattice knot of type
$3_1^+$ is instead considered but in the more tightly containing
average excluded volume $V_e$.  In this case the partition function
is given by
\begin{eqnarray}
\C{Z}_{SC} (3_1^+) 
&=&84e^{-32p/3}+24e^{-65p/6}+300e^{-10p}+96e^{-61p/6} \nonumber \\
& &+240e^{-59p/6}+128e^{-21p/2}+48{{e}^{-28p/3}}+168{{e}^{-31p/3}}\nonumber \\
& &+48e^{-19/2p}+72e^{-29p/3}+12e^{-11p} +72e^{-103p/12} \nonumber \\
& &+192e^{-53p/6}+48e^{-26p/3}+24e^{-35p/4}+48e^{-17/2p} \nonumber  \\
& &+24e^{-25p/3}+24e^{-101p/12}+12e^{-8p} .
\end{eqnarray}
The compressibility can be determined in this model as above.  
It is plotted in figure \ref{FIG3}.
%%%%%%%%%%%%%%%%%%%%%%%%%%%%%%%%%%%%%%%%%%%%%%%%
%%%%%%%%%%%%%%%%%%%%%%%%%%%%%%%%%%%%%%%%%%%%%%%%%%%%%%
In this case one notes that the scale of the $\beta$-axis is
an order of magnitude smaller than the case in figure \ref{FIG2}.
The pressure range on the $X$-axis is also expanded to larger
pressure (up to $p=16$ in this case, compared to $p=4$ in 
figure \ref{FIG2}).  The qualitative shape of the curve is also
slightly changed; there is still a global maximum at a non-zero
value of $p$, but the secondary local maximum has disappeared.
Nevertheless, the general shape of the curve is comparable to
the curve in figure \ref{FIG2}.  The global maximum is at
$p=0.61046\ldots$ and where $\beta=0.05907\ldots$.

The maximum amount of useful work that can be performed by
letting a lattice knot type $3_1^+$ expand from its minimal 
containing volume to its equilibrium at zero pressure can be
computed from the free energy of the model.  If the volume $V_b$
is used, then $\C{F}_{SC} (3_1^+) = - \log \C{Z}_{SC} (3_1^+)$
with the partition function given by equation \Ref{eqnZSC31}.

By identifying the smallest value of $V_b$ in equation \Ref{eqnZSC31},
putting $p=0$ and determining the free energy difference between
the compressed and zero pressure state, it follows that the
maximum amount of useful work is $\C{W}_{3_1^+} = \log 1664 - \log 12
= \log (416/3) = 4.93207\ldots$.  Interestingly, in this case 
the same result is obtained if one uses the average excluded 
volume instead.

\subsubsection{Compressibility of $4_1$:}
\begin{figure}[h!]
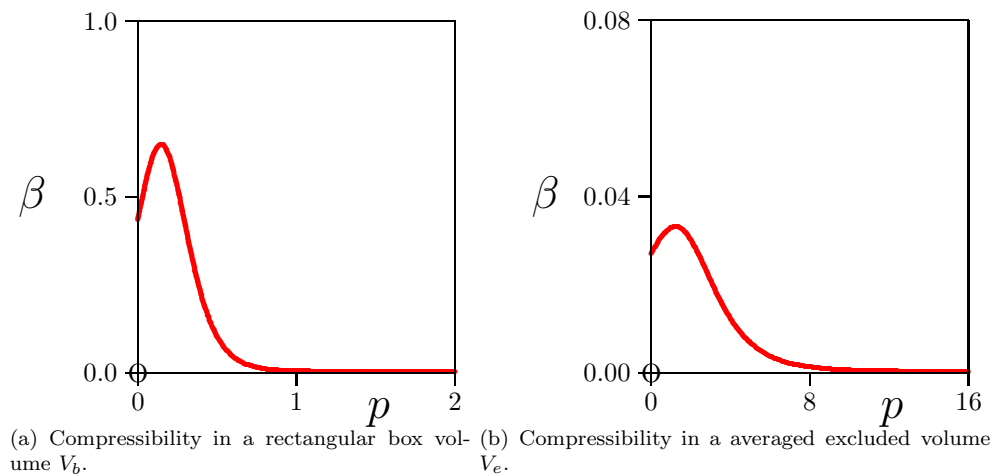

\centering
\subfigure[Compressibility in a rectangular box volume $V_b$.]{
\input{figure4A}
\label{FIG4A}}
\subfigure[Compressibility in a averaged excluded volume $V_e$.]{
\input{figure4B}
\label{FIG4B}}
\caption{Compressibility of the minimal lattice knots of type $4_1$
in the SC lattice.}
\label{FIG4} %ZXZ[FIG4]
\end{figure}

The compressibility of the figure eight knot $4_1$ is illustrated
in figure \ref{FIG4}:  In figure \ref{FIG4A} the compressibility
in a rectangular box volume $V_b$ is shown, and in figure
\ref{FIG4B} the compressibility in an averaged excluded volume $V_e$
is illustrated.  Note the different scales on the axes of the two
graphs.
%%%%%%%%%%%%%%%%%%%%%%%%%%%%%%%%%%%%%%%%%%%%%%%%
%%%%%%%%%%%%%%%%%%%%%%%%%%%%%%%%%%%%%%%%%%%%%%%%%%%%%%
One may similarly determine the maximum amount of useful work
if the lattice knot is relaxed from its smallest containing
volume to its state at zero pressure.  If the volume $V_b$ is
used, then this is $\C{W}_{4_1} = \log(38/9) = 1.44036\ldots$,
and if $V_e$ is used, then $\C{W}_{4_1} = \log 38 = 3.63759\ldots$.

\subsubsection{Compressibility of $5_1^+$ and $5_2^+$:}
\begin{figure}[h!]
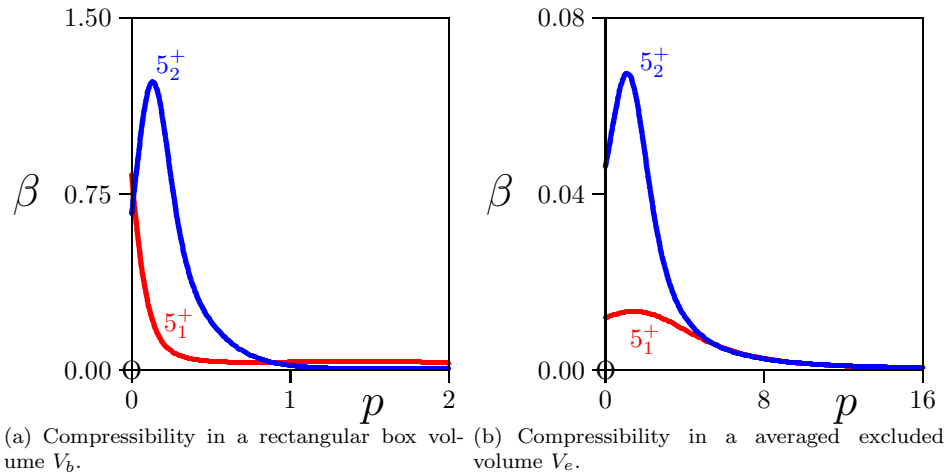

\centering
\subfigure[Compressibility in a rectangular box volume $V_b$.]{
\input{figure5A}
\label{FIG5A}}
\subfigure[Compressibility in a averaged excluded volume $V_e$.]{
\input{figure5B}
\label{FIG5B}}
\caption{Compressibility of the minimal lattice knots of types $5_1^+$
and $5_2^+$ in the SC lattice.}
\label{FIG5} %ZXZ[FIG5]
\end{figure}

The compressibilities of the two knots $5_1^+$ and $5_2^+$ are
plotted in figure \ref{FIG5}.  For both choices of the volumes,
the knot $5_2^+$ is more compressible than $5_1^+$.  The generic
shapes of the compressibility of $5_2^+$ are similar for the
two volumes, both cases having a global maximum away from zero pressure. 
%%%%%%%%%%%%%%%%%%%%%%%%%%%%%%%%%%%%%%%%%%%%%%%%
%%%%%%%%%%%%%%%%%%%%%%%%%%%%%%%%%%%%%%%%%%%%%%%%%%%%%%
In the case of $5_1^+$ on the other hand, the global maximum in the
compressibility is at zero pressure when the volume $V_b$ is used,
but it is at $p>0$ when $V_e$ is used.  The maxima in figure
\ref{FIG5A} is at $p=0$ ($\beta=0.82695\ldots$) for $5_1^+$
and $p=0.13237\ldots$ ($\beta=1.22294\ldots$) for $5_2^+$.  Observe
that $5_1^+$ has a local maximum at $p=1.43622\ldots$ 
($\beta=0.03260\ldots$).  In figure \ref{FIG5B} the global
maxima are located at $p=1.43811\ldots$ ($\beta=0.01308\ldots$)
and $p=1.11177\ldots$ ($\beta=0.06718\ldots$) respectively.

The maximum amount of work that can be extracted from these
two knot types were, for $5_1^+$, $\C{W}_{5_1^+} = \log (139/5) = 3.32504\ldots$
with volume $V_b$ and $\C{W}_{5_1^+} = \log 139 = 4.93447\ldots$ with
volume $V_e$, and for $5_2^+$, $\C{W}_{5_2^+} = \log (133/5) 
= 3.28091\ldots$ with volume $V_b$ and $\C{W}_{5_2^+} = \log 2394
= 7.78072\ldots$ with volume $V_e$.

\subsubsection{Compressibility of knots with 6 crossings:}
\begin{figure}[h!]
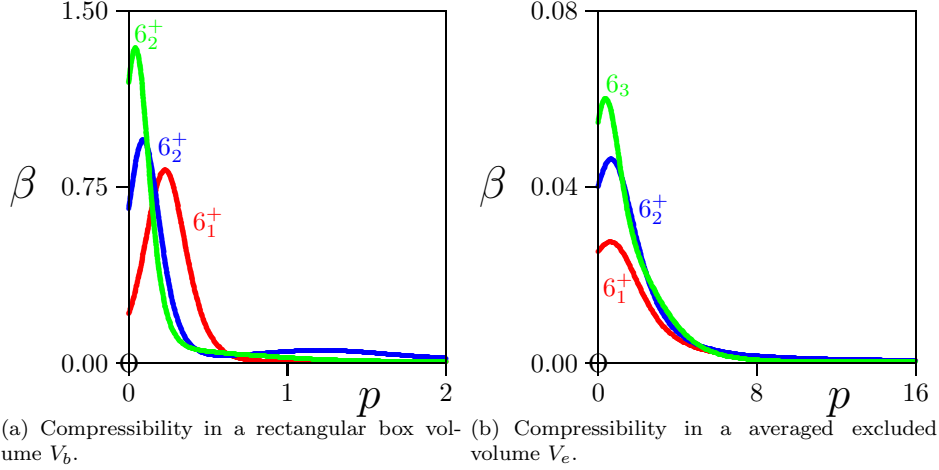

\centering
\subfigure[Compressibility in a rectangular box volume $V_b$.]{
\input{figure6A}
\label{FIG6A}}
\subfigure[Compressibility in a averaged excluded volume $V_e$.]{
\input{figure6B}
\label{FIG6B}}
\caption{Compressibility of the minimal lattice knots of types $6_1^+$,
$6_2^+$ and $6_3$ in the SC lattice.}
\label{FIG6} %ZXZ[FIG6]
\end{figure}

The compressibility of six crossing knots are displayed in figure
\ref{FIG6}.  For both choices of $V_b$ and $V_e$ as containing volumes,
the knot $6_3$ were more compressible, and $6_1^+$ least compressible.

In all three cases the maximal compressiblity were at non-zero pressure,
and only in the case of $6_2^+$ with volume $V_b$ did we observe a local
maximum in the compressibility at higher pressure, compared to the global
maximum at low pressure. By using the volume $V_b$ to measure compressibility,
%%%%%%%%%%%%%%%%%%%%%%%%%%%%%%%%%%%%%%%%%%%%%%%%
%%%%%%%%%%%%%%%%%%%%%%%%%%%%%%%%%%%%%%%%%%%%%%%%%%%%%%
the (global) maxima in the compressibility were at $p=0.22958\ldots$
where $\beta=0.81816\ldots$ for $6_1^+$, at $p=0.09103\ldots$ 
where $\beta=0.94524\ldots$ for $6_2^+$ (with a local maximum at $p=1.20977\ldots$
where $\beta=0.04840\ldots$), and at $p=0.04118\ldots$ where
$\beta=1.33719\ldots$ for $6_3$.

The maximum amount for work that can be extracted from these knot types
are $\C{W}_{6_1^+} = \log (128/9) = 2.65481\ldots$, 
$\C{W}_{6_2^+} = \log 171 = 5.14166\ldots$, 
and $\C{W}_{6_3} = \log (74/9) = 2.10684\ldots$. 

Using the volume $V_e$ instead, one finds that global maxima at
$p=0.61767\ldots$ where $\beta=0.02723\ldots$ for $6_1^+$, at $p=0.66097\ldots$ 
where $\beta=0.04609\ldots$ for $6_2^+$, and at $p=0.37398\ldots$ where
$\beta=0.05988\ldots$ for $6_3$.

The maximum amount for work that can be extracted from these knot types
are $\C{W}_{6_1^+} = \log 256 = 5.54517\ldots$, 
$\C{W}_{6_2^+} = \log (36/19) = 6.52795\ldots$, 
and $\C{W}_{6_3} = \log 74 = 4.30406\ldots$. 

\subsubsection{Compressibility of knots with 7 or more crossings:}
The compressibilities of 7 and 8 crossings knots for the choice $V_b$ are
summarized in table \ref{CompKnot-Box}.
%%%%%%%%%%%%%%%%%%%%%%%%%%%%%%%%%%%%%%%%%%%%%%%%%%%%%%%%%%%%%%%%%%%%%%%%%%%%%%%
%%%%%%%%%%%%%%%%%%%%%%%%%%%%%%%%%%%%%%%%%%%%%%%%%%%%%%%%%%%%%%%%%%%%%%%%%%%%%%%
\begin{table}[h!]
\begin{center}
 \begin{tabular}{||c||c||c|c||c|c||r||r||}
 \hline
 \fns Knot & \fns $\beta(0)$& \fns $p_m$  & \fns $\max\beta(p)$ & \fns loc $p_m$ & \fns local $\max\beta(p)$ & \fns $\C{W}_K$ & \fns Min $V_b$  \\
  \hline  
$0_1$   &\fns $0$ &\fns $-$ &\fns $-$ &\fns $-$ &\fns $-$ & \fns $0$ & \fns $0$ \\
$3_1^+$ &\fns $0.65347$ &\fns $0.13557$ &\fns $0.92845$ &\fns $1.82115$ &\fns $0.05888$  &\fns $4.93207$ &\fns $16$ \\
$4_1$   &\fns $0.43228$ &\fns $0.14599$ &\fns $0.64617$ &\fns $-$ &\fns $-$              &\fns $1.44036$ &\fns $27$ \\
$5_1^+$ &\fns $0.82695$ &\fns $0$       &\fns $0.82695$ &\fns $1.43622$ &\fns $0.03260$  &\fns $3.32504$ &\fns $30$ \\
$5_2^+$ &\fns $0.66116$ &\fns $0.13237$ &\fns $1.22294$ &\fns $-$ &\fns $-$              &\fns $3.28091$ &\fns $30$ \\
$6_1^+$ &\fns $0.20546$ &\fns $0.22958$ &\fns $0.81816$ &\fns $-$ &\fns $-$              &\fns $2.65481$ &\fns $36$ \\
$6_2^+$ &\fns $0.65497$ &\fns $0.09103$ &\fns $0.94525$ &\fns $1.20977$ &\fns $0.04840$  &\fns $5.14166$ &\fns $45$ \\
$6_3$   &\fns $1.18846$ &\fns $0.04119$ &\fns $1.33719$ &\fns $-$ &\fns $-$              &\fns $2.10684$ &\fns $45$ \\
$7_1^+$ &\fns $1.43410$ &\fns $0.06558$ &\fns $1.93627$ &\fns $-$ &\fns $-$              &\fns $1.55110$ &\fns $40$ \\
$7_2^+$ &\fns $1.56522$ &\fns $0$       &\fns $1.56522$ &\fns $-$ &\fns $-$              &\fns $2.76117$ &\fns $45$ \\
$7_3^+$ &\fns $0$       &\fns $-$       &\fns $-$       &\fns $-$ &\fns $-$              &\fns $0$ &\fns $45$ \\
$7_4^+$ &\fns $0.05340$ &\fns $0$       &\fns $0.05340$ &\fns $-$ &\fns $-$              &\fns $0.33647$ &\fns $60$ \\
$7_5^+$ &\fns $0.11667$ &\fns $0.23179$ &\fns $1.28010$ &\fns $-$ &\fns $-$              &\fns $5.28320$ &\fns $45$ \\
$7_6^+$ &\fns $0.97333$ &\fns $0.03759$ &\fns $1.07580$ &\fns $-$ &\fns $-$              &\fns $2.18183$ &\fns $60$ \\
$7_7^+$ &\fns $0.76190$ &\fns $0.05029$ &\fns $0.89165$ &\fns $-$ &\fns $-$              &\fns $1.09861$ &\fns $64$ \\
$8_1^+$ &\fns $1.31169$ &\fns $0.04611$ &\fns $1.49428$ &\fns $-$ &\fns $-$              &\fns $3.56449$ &\fns $48$ \\
$8_2^+$ &\fns $2.49340$ &\fns $0$       &\fns $2.49340$ &\fns $-$ &\fns $-$              &\fns $4.61042$ &\fns $48$ \\
$8_3$ &\fns $3.98158$ &\fns $0$       &\fns $3.98158$ &\fns $-$ &\fns $-$              &\fns $2.12193$ &\fns $27$ \\
$8_4^+$ &\fns $0.23441$ &\fns $0.18360$ &\fns $1.03728$ &\fns $-$ &\fns $-$              &\fns $3.68587$ &\fns $60$ \\
$8_5^+$ &\fns $0$ &\fns $-$ &\fns $-$ &\fns $-$ &\fns $-$                                &\fns $0$ &\fns $80$ \\
$8_6^+$ &\fns $0$ &\fns $-$ &\fns $-$ &\fns $-$ &\fns $-$                                &\fns $0$ &\fns $80$ \\
$8_7^+$ &\fns $0$ &\fns $-$ &\fns $-$ &\fns $-$ &\fns $-$                                &\fns $0$ &\fns $60$ \\
$8_8^+$ &\fns $0$ &\fns $-$ &\fns $-$ &\fns $-$ &\fns $-$                                &\fns $0$ &\fns $80$ \\
$8_9$ &\fns $0.60841$ &\fns $0.09961$ &\fns $1.43594$ &\fns $-$ &\fns $-$              &\fns $1.99470$ &\fns $60$ \\
$8_{10}^+$ &\fns $0$ &\fns $-$ &\fns $-$   &\fns $-$ &\fns $-$                           &\fns $0$ &\fns $80$ \\
$8_{11}^+$ &\fns $0$ &\fns $-$ &\fns $-$   &\fns $-$ &\fns $-$                           &\fns $0$ &\fns $80$ \\
$8_{12}$ &\fns $1.21212$ &\fns $0.04185$ &\fns $1.43594$ &\fns $-$ &\fns $-$           &\fns $1.09861$ &\fns $60$ \\
$8_{13}^+$ &\fns $1.78904$ &\fns $0$       &\fns $1.78904$ &\fns $-$ &\fns $-$           &\fns $1.62612$ &\fns $64$ \\
$8_{14}^+$ &\fns $0$ &\fns $-$ &\fns $-$   &\fns $-$ &\fns $-$                           &\fns $0$ &\fns $80$ \\
$8_{15}^+$ &\fns $2.36553$ &\fns $0$       &\fns $2.36553$ &\fns $-$ &\fns $-$           &\fns $4.45076$ &\fns $48$ \\
$8_{16}^+$ &\fns $0$ &\fns $-$ &\fns $-$ &\fns $-$ &\fns $-$                             &\fns $0$ &\fns $80$ \\
$8_{17}$ &\fns $1.26959$ &\fns $0.04260$ &\fns $1.43108$ &\fns $-$ &\fns $-$           &\fns $3.54458$ &\fns $64$ \\
$8_{18}$ &\fns $0$ &\fns $-$ &\fns $-$ &\fns $-$ &\fns $-$                             &\fns $0$ &\fns $100$ \\
$8_{19}^+$ &\fns $1.16978$ &\fns $0.04525$ &\fns $1.34959$ &\fns $-$ &\fns $-$           &\fns $2.09152$ &\fns $45$ \\
$8_{20}^+$ &\fns $0$ &\fns $-$ &\fns $-$ &\fns $-$ &\fns $-$                             &\fns $0$ &\fns $60$ \\
$8_{21}^+$ &\fns $0.36279$ &\fns $0.15514$ &\fns $0.95539$ &\fns $-$ &\fns $-$           &\fns $2.33082$ &\fns $48$ \\
$9_1^+$    &\fns $2.99129$ &\fns $0$ &\fns $2.99129$ &\fns $-$ &\fns $-$                 &\fns $1.62095$ &\fns $50$ \\
$9_2^+$    &\fns $1.97047$ &\fns $0$ &\fns $1.97047$ &\fns $0.40139$ &\fns $0.60045$     &\fns $7.39488$ &\fns $48$ \\
$9_{42}^+$ &\fns $0.80392$ &\fns $0$ &\fns $0.80392$ &\fns $-$ &\fns $-$                 &\fns $2.00459$ &\fns $60$ \\
$9_{47}^+$ &\fns $1.81376$ &\fns $0$ &\fns $1.81376$ &\fns $-$ &\fns $-$                 &\fns $0.26236$ &\fns $80$ \\
$10_1^+$   &\fns $1.66066$ &\fns $0$ &\fns $1.66066$ &\fns $0.18995$ &\fns $0.98320$     &\fns $6.97624$ &\fns $54$ \\
$10_2^+$   &\fns $3.12685$ &\fns $0$ &\fns $3.12685$ &\fns $-$ &\fns $-$                 &\fns $4.57009$ &\fns $60$ \\
  \hline
 \end{tabular}
\end{center}
 \caption{Compressibility of SC Lattice Knots with volume $V_b$.}
  \label{CompKnot-Box} %ZXZ[CompKnot-Box] 
\vspace{1mm}
\end{table} 
%%%%%%%%%%%%%%%%%%%%%%%%%%%%%%%%%%%%%%%%%%%%%%%%%%%%%%%%%%%%%%%%%%%%%%%%%%%%%%%
%%%%%%%%%%%%%%%%%%%%%%%%%%%%%%%%%%%%%%%%%%%%%%%%%%%%%%%%%%%%%%%%%%%%%%%%%%%%%%%
Several minimal lattice knots in table \ref{CompKnot-Box} are incompressible,
for example $0_1$, $7_3^+$, $8_5^+$, $8_6^+$ and so on.  The largest compressibility
at zero pressure is for the knot $8_3$, and this is also the largest 
compressibility overall in the table.  In this table the appearance of
a local maximum in the compressibility, as seen for the trefoil in figure
\ref{FIG2}, for example, appears to be the exception, rather than the
rule. Only the knot types $3_1^+$, $5_1^+$, $6_2^+$, $9_2^+$ and $10_1^+$
have local maxima in their compressibility when pressurized in 
a rectangular box.  Most knot types have maximum useful work between 
$0$ and $3$ in lattice units, but for the knot types 
$3_1^+$, $6_2^+$, $7_5^+$, $8_2^+$, $8_{15}^+$, $9_2^+$, $10_1^+$ 
and $10_2^+$ this exceeds $4$ lattice units.

The minimal value of $V_b$ is listed in the last column of table \ref{CompKnot-Box}.
This shows for example that a trefoil can be tied in the cubic lattice in
a rectangular box of volume $16$.  Comparison of the results in this list shows
that amongst 6 crossings knots it is $6_1^+$ which can be tied in the smallest
box, and amongst 7 crossings knots $7_1^+$.  In the list of 8 crossings prime
knots, type $8_3$ can be tied in a box of volume $27$, far smaller than any
other 8 crossing knot (or even any 5, 6 or 7 crossing knot).  This suggests
that $8_3$ can be embedded very efficiently, using relatively little volume,
in the SC lattice.

\begin{figure}[h!]
\centering
\input{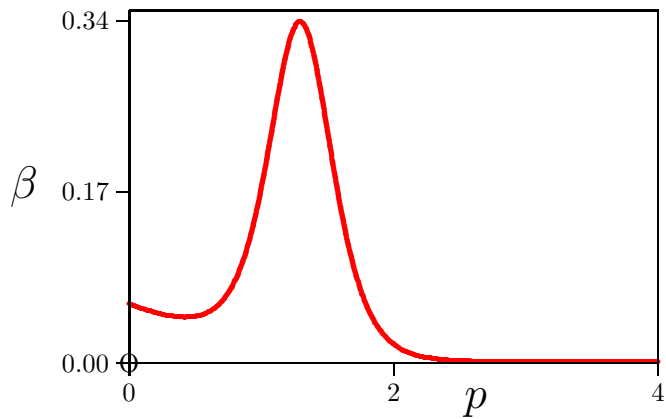}
\caption{Compressibility of the SC minimal lattice knot of
type $10_1^+$ in its averaged excluded volume $V_e$.}
\label{FIG7} %ZXZ[FIG7]
\end{figure}

In table \ref{CompKnot-Box-CH} the compressibilities of minimal SC lattices knots
with the choice of the volume $V_e$ are displayed.  Generally, the compressibilities
are smaller than those in table \ref{CompKnot-Box} -- this is to be expected
because the average excluded volume $V_e$ is a more tightly fitting volume
about the lattice knots, compared to the rectangular box volume $V_b$.
%%%%%%%%%%%%%%%%%%%%%%%%%%%%%%%%%%%%%%%%%%%%%%%%%%%%%%%%%%%%%%%%%%%%%%%%%%%%%%%%%
%%%%%%%%%%%%%%%%%%%%%%%%%%%%%%%%%%%%%%%%%%%%%%%%%%%%%%%%%%%%%%%%%%%%%%%%%%%%%%%%%
\begin{table}[h!]
\begin{center}
 \begin{tabular}{||c||c||c|c||c|c||r||r||}
 \hline
 \fns Knot &\fns  $\beta(0)$&\fns  $p_m$  &\fns  $\max\beta(p)$ & \fns loc $p_m$ & \fns local $\max\beta(p)$ &\fns  $\C{W}_K$ &\fns  Min $V_e$ \\
  \hline  
$0_1$   &\fns $0$ &\fns $-$ &\fns $-$ &\fns $-$ &\fns $-$              &\fns $0$ &\fns $0$ \\
$3_1^+$ &\fns $0.05167$ &\fns $0.61046$ &\fns $0.05907$ &\fns $-$ &\fns $-$              &\fns $4.93207$ &\fns $8$\\
$4_1$   &\fns $0.02684$ &\fns $1.22731$ &\fns $0.03290$ &\fns $-$ &\fns $-$              &\fns $3.63759$ &\fns $12$ \\
$5_1^+$ &\fns $0.01159$ &\fns $1.43811$ &\fns $0.01308$ &\fns $-$ &\fns $-$              &\fns $4.93447$ &\fns $14$ \\
$5_2^+$ &\fns $0.04600$ &\fns $1.11177$ &\fns $0.06718$ &\fns $-$ &\fns $-$              &\fns $7.78072$ &\fns $44/3$ \\
$6_1^+$ &\fns $0.02500$ &\fns $0.61768$ &\fns $0.02724$ &\fns $-$ &\fns $-$              &\fns $5.54518$ &\fns $37/2$ \\
$6_2^+$ &\fns $0.03982$ &\fns $0.66098$ &\fns $0.04609$ &\fns $-$ &\fns $-$              &\fns $6.52796$ &\fns $113/6$ \\
$6_3$   &\fns $0.05438$ &\fns $0.37398$ &\fns $0.05989$ &\fns $-$ &\fns $-$              &\fns $4.30407$ &\fns $37/2$ \\
$7_1^+$ &\fns $0.03748$ &\fns $0.15097$ &\fns $0.03787$ &\fns $-$ &\fns $-$              &\fns $6.56174$ &\fns $121/6$ \\
$7_2^+$ &\fns $0.05004$ &\fns $0.02186$ &\fns $0.05005$ &\fns $-$ &\fns $-$              &\fns $8.85474$&\fns $85/4$ \\
$7_3^+$ &\fns $0.00354$ &\fns $0.53600$ &\fns $0.00357$ &\fns $-$ &\fns $-$              &\fns $2.30259$ &\fns $65/3$ \\
$7_4^+$ &\fns $0.00706$ &\fns $0.93823$ &\fns $0.00819$ &\fns $-$ &\fns $-$              &\fns $1.25276$ &\fns $68/3$ \\
$7_5^+$ &\fns $0.01311$ &\fns $1.95942$ &\fns $0.04274$ &\fns $-$ &\fns $-$              &\fns $5.28320$ &\fns $68/3$ \\
$7_6^+$ &\fns $0.03711$ &\fns $0$       &\fns $0.03711$ &\fns $-$ &\fns $-$              &\fns $5.87071$ &\fns $143/6$ \\
$7_7^+$ &\fns $0.03062$ &\fns $0.56027$ &\fns $0.03859$ &\fns $-$ &\fns $-$              &\fns $1.65823$ &\fns $139/6$ \\
$8_1^+$ &\fns $0.06928$ &\fns $0$       &\fns $0.06928$ &\fns $-$ &\fns $-$              &\fns $6.20355$ &\fns $73/3$ \\
$8_2^+$ &\fns $0.07573$ &\fns $0$       &\fns $0.07573$ &\fns $-$ &\fns $-$              &\fns $7.55486$ &\fns $293/12$ \\ 
$8_3$   &\fns $0.02684$ &\fns $1.22731$ &\fns $0.03290$ &\fns $-$ &\fns $-$              &\fns $3.63759$ &\fns $12$ \\ 
$8_4^+$ &\fns $0.03488$ &\fns $0.53631$ &\fns $0.03840$ &\fns $-$ &\fns $-$              &\fns $6.21160$ &\fns $311/12$ \\
$8_5^+$ &\fns $0.01102$ &\fns $0.08509$ &\fns $0.01103$ &\fns $-$ &\fns $-$              &\fns $3.17805$ &\fns $167/6$ \\ 
$8_6^+$ &\fns $0.01169$ &\fns $0.63470$ &\fns $0.01226$ &\fns $-$ &\fns $-$              &\fns $5.43808$ &\fns $167/6$ \\
$8_7^+$ &\fns $0$       & $-$           & $-$           &\fns $-$ &\fns $-$              &\fns $0$ &\fns $77/3$ \\
$8_8^+$ &\fns $0.01758$ &\fns $0$       &\fns $0.01758$ &\fns $-$ &\fns $-$              &\fns $3.48124$ &\fns $57/2$ \\
$8_9$   &\fns $0.04620$ &\fns $0.55186$ &\fns $0.05906$ &\fns $-$ &\fns $-$              &\fns $6.59987$ &\fns $79/3$ \\
$8_{10}^+$ &\fns $0.00935$ &\fns $0.47816$ &\fns $0.00954$ &\fns $-$ &\fns $-$              &\fns $3.55535$ &\fns $28$ \\
$8_{11}^+$ &\fns $0.00299$ &\fns $0.01757$ &\fns $0.00299$ &\fns $-$ &\fns $-$              &\fns $0.69315$ &\fns $169/6$ \\
$8_{12}$   &\fns $0.01639$ &\fns $0.47743$ &\fns $0.01709$ &\fns $-$ &\fns $-$              &\fns $3.98898$ &\fns $83/3$ \\ 
$8_{13}^+$ &\fns $0.05487$ &\fns $0.17977$ &\fns $0.05581$ &\fns $-$ &\fns $-$              &\fns $6.29895$ &\fns $161/6$ \\ 
$8_{14}^+$ &\fns $0.00514$ &\fns $0.23152$ &\fns $0.00516$ &\fns $-$ &\fns $-$              &\fns $2.70805$ &\fns $169/6$ \\ 
$8_{15}^+$ &\fns $0.05797$ &\fns $0.77810$ &\fns $0.07990$ &\fns $-$ &\fns $-$              &\fns $7.42118$ &\fns $76/3$ \\
$8_{16}^+$ &\fns $0.00367$ &\fns $0.01653$ &\fns $0.00367$ &\fns $-$ &\fns $-$              &\fns $0.69315$ &\fns $359/12$ \\ 
$8_{17}$   &\fns $0.03236$ &\fns $1.33662$ &\fns $0.03686$ &\fns $-$ &\fns $-$              &\fns $5.62402$ &\fns $343/12$ \\
$8_{18}$   &\fns $0.01720$ &\fns $0.53164$ &\fns $0.01797$ &\fns $-$ &\fns $-$              &\fns $3.61092$ &\fns $377/12$ \\ 
$8_{19}^+$ &\fns $0.06651$ &\fns $0.47415$ &\fns $0.08667$ &\fns $-$ &\fns $-$              &\fns $6.36819$ &\fns $107/6$ \\ 
$8_{20}^+$ &\fns $0.00291$ &\fns $0.60279$ &\fns $0.00298$ &\fns $-$ &\fns $-$              &\fns $1.60944$ &\fns $275/12$ \\
$8_{21}^+$ &\fns $0.05177$ &\fns $0.54428$ &\fns $0.05510$ &\fns $-$ &\fns $-$              &\fns $7.06262$ &\fns $263/12$ \\
$9_{1}^+$  &\fns $0.07013$ &\fns $0.31127$ &\fns $0.07755$ &\fns $-$ &\fns $-$              &\fns $8.88288$ &\fns $103/4$ \\
$9_{2}^+$  &\fns $0.04825$ &\fns $0.66653$ &\fns $0.05275$ &\fns $-$ &\fns $-$              &\fns $11.13255$ &\fns $169/6$ \\
$9_{42}^+$ &\fns $0.02316$ &\fns $0.12142$ &\fns $0.02312$ &\fns $-$ &\fns $-$              &\fns $6.36130$ &\fns $289/12$ \\
$9_{47}^+$ &\fns $0.02259$ &\fns $1.15783$ &\fns $0.19259$ &\fns $-$ &\fns $-$              &\fns $5.65599$ &\fns $101/4$ \\
$10_{1}^+$ &\fns $0.05728$ &\fns $1.29256$ &\fns $0.33799$ &\fns $-$ &\fns $-$              &\fns $9.17347$ &\fns $83/3$ \\
$10_{2}^+$ &\fns $0.07893$ &\fns $0$       &\fns $0.07893$ &\fns $-$ &\fns $-$              &\fns $9.11339$ &\fns $377/12$ \\
\hline
 \end{tabular}
\end{center}
 \caption{Compressibility of SC Lattice Knots with volume $V_e$.}
  \label{CompKnot-Box-CH} %ZXZ[CompKnot-Box-CH]
\vspace{1mm}
\end{table} 
%%%%%%%%%%%%%%%%%%%%%%%%%%%%%%%%%%%%%%%%%%%%%%%%%%%%%%%%%%%%%%%%%%%%%%%%%%%%%%%
%%%%%%%%%%%%%%%%%%%%%%%%%%%%%%%%%%%%%%%%%%%%%%%%%%%%%%%%%%%%%%%%%%%%%%%%%%%%%%%
All the minimal lattice knots in table \ref{CompKnot-Box-CH} are 
compressible, with the exception of the unknot and the knot type $8_7^+$.  
The largest compressibility is obtained for knot type $10_{1}^+$ at 
$p=1.29256\ldots$ where $\beta=0.33799\ldots$.  In knot types such as $10_1^+$,
the compressibility varies dramatically with pressure, and this example is
illustrated in figure \ref{FIG7}.  A similar profile can be plotted for $9_{47}^+$.
The largest compressibility at zero pressure is found for knot type $10_2^+$.
%%%%%%%%%%%%%%%%%%%%%%%%%%%%%%%%%%%%%%%%%%%%%%%%
%%%%%%%%%%%%%%%%%%%%%%%%%%%%%%%%%%%%%%%%%%%%%%%%%%%%%%
Observe that there are no knot types with a secondary local maximum in
the compressibility in table \ref{CompKnot-Box-CH}. Most knot types have maximum 
useful work between $0$ and $6$ in lattice units, but for the knot types 
$5_2^+$, $6_2^+$, $7_1^+$, $7_2^+$, $8_1^+$, $8_2^+$, $8_4^+$, $8_9$,
$8_{13}^+$, $8_{15}^+$, $8_{19}^+$, $8_{21}^+$, $9_1^+$, $9_2^+$, $10_1^+$ 
and $10_2^+$ this exceeds $6$ lattice units.  The knot type $9_2^+$ has
particularly large value for the work, namely $\C{W}_{9_2^+} = 11.13255\ldots$
far larger than types $10_1^+$ and $10_2^+$, where the work exceeds $9$
units. 

The minimal value of $V_e$ is listed in the last column of table
\ref{CompKnot-Box-CH}. The minimal value for $3_1^+$ is $8$, which is
only one half of the volume of the minimal size rectangular box containing
a lattice knot of type $3_1^+$.  This shows that the volume $V_e$ is far 
smaller than $V_b$ for this knot type, and the averaged excluded
volume of the knot type is a far tighter fit about the lattice knot.
Comparison of the results in this list shows similarly that $8_3$ can
be tied with relatively very small value of $V_e$, in this case $V_e=12$,
smaller than the minimal value of $V_e$ for any other $5$, $6$, $7$ or
$8$ crossing minimal lattice knots.

\subsection{Compressibility of minimal lattice knots in the FCC and BCC lattices}

The compressibility of minimal lattice knots in the FCC lattice can similarly
be computed, using again two choices of an enclosing volume, namely $V_b$ for
the volume of the minimal rectangular box containing the polygon, and $V_e$
for the volume of the average excluded volume.  

The partition functions of FCC minimal lattice knots up to crossing number $7$
and with volumes $V_b$ are listed in table \ref{FCC-PartKnot-Box}.
%%%%%%%%%%%%%%%%%%%%%%%%%%%%%%%%%%%%%%%%%%%%%%%%%%%%%%%%%%%%%%%%%%%%%%%%%%%%%%%%%
\begin{table}[h!]
\begin{center}
 \begin{tabular}{||c||l||}
 \hline
  Knot & \multicolumn{1}{|c||}{$Z_n(p)$} \\
  \hline  
$0_1$   &\fns $8$ \\
$3_1^+$ &\fns $16{e^{-27p}}+48{e^{-24p}}$ \\
$4_1$   &\fns $24{e^{-64p}}+576{e^{-48p}}+348{e^{-45p}}+1416{e^{-36p}}+192{e^{-32p}}+48{e^{-30p}}+192{e^{-27p}}$ \\
$5_1^+$ &\fns $48{e^{-48p}}+24{e^{-45p}}+24{e^{-36p}}$ \\
$5_2^+$ &\fns $48{e^{-80p}}+96{e^{-64p}}+192{e^{-60p}}+288{e^{-48p}}+96{e^{-45p}}+48{e^{-40p}} $ \\
$6_1^+$ &\fns $48{e^{-125p}}+816{e^{-100p}}+144{e^{-96p}}+4224{e^{-80p}}+1104{e^{-75p}}+912{e^{-72p}}$ \\
        &\fns $+2448{e^{-64p}}+5712{e^{-60p}}+816{
e^{-54p}}+2112{e^{-48p}}+624{e^{-45p}}+48{e^{-40p}}$ \\
$6_2^+$ &\fns $960{e^{-80p}}+288{e^{-75p}}+48{e^{-72p}}+864{e^{-64p}}+1344{e^{-60p}}+1152{e^{-48p}}+384{e^{-45p}}$ \\
$6_3$   &\fns $960{e^{-100p}}+48{e^{-96p}}+18336{e^{-80p}}+4896{e^{-75p}}+1440{e^{-72p}}+24096{e^{-64p}}+24864{e^{-60p}}$ \\
        &\fns $+96{e^{-54p}}+23280{e^{-48p}}+3840{e^{-45p}}+864{e^{-36p}}$ \\
$7_1^+$ &\fns $336{e^{-100p}}+2400{e^{-80p}}+144{e^{-75p}}+96{e^{-72p}}+192{e^{-64p}}+864{e^{-60p}}+48{e^{-54p}}$ \\
$7_2^+$ &\fns $48{e^{-120p}}+432{e^{-100p}}+96{e^{-96p}}+96{e^{-90p}}+1200{e^{-80p}}+552{e^{-75p}}+432{e^{-72p}}+336{e^{-64p}}$ \\
        &\fns $+672{e^{-60p}}+168{e^{-54p}}+96{e^{-45p}}$ \\
$7_3^+$ &\fns $96{e^{-100p}}+96{e^{-96p}}+288{e^{-80p}}+96{e^{-75p}}+144{e^{-72p}}+48{e^{-64p}}+192{e^{-60p}}$ \\
$7_4^+$ &\fns $72{e^{-100p}}+24{e^{-96p}}$ \\
$7_5^+$ &\fns $144{e^{-125p}}+624{e^{-120p}}+6192{e^{-100p}}+1920{e^{-96p}}+720{e^{-90p}}+8112{e^{-80p}}+2544{e^{-75p}}$ \\
        &\fns $+1488{e^{-72p}}+2064{e^{-64p}}+3216{e^{-60p}}+432{e^{-54p}}$ \\
$7_6^+$ &\fns $480{e^{-100p}}+1728{e^{-80p}}+96{e^{-75p}}+1728{e^{-64p}}+288{e^{-60p}}+576{e^{-48p}}$ \\
$7_7^+$ &\fns $96{e^{-100p}}+384{e^{-80p}}+48{e^{-72p}}+384{e^{-64p}}+192{e^{-60p}}+192{e^{-48p}}$ \\
 \hline
 \end{tabular}
\end{center}
 \caption{Partition Functions: Minimal FCC lattice knots pressurized in a rectangular box.}
  \label{FCC-PartKnot-Box}
\end{table}
%%%%%%%%%%%%%%%%%%%%%%%%%%%%%%%%%%%%%%%%%%%%%%%%%%%%%%%%%%%%%%%%%%%%%%%%%%%%%%%%%
%%%%%%%%%%%%%%%%%%%%%%%%%%%%%%%%%%%%%%%%%%%%%%%%%%%%%%%%%%%%%%%%%%%%%%%%%%%%%%%%%
With the choice of the average excluded volume the partition functions are
similarly more complicated, and we do not reproduce those here.

The partition function of minimal lattice knots in the BCC lattice with the
volume $V_b$ is similarly given in table \ref{BCC-PartKnot-Box}.  Far more lengthy
expressions are obtained when the average excluded volume $V_e$ is used.
%%%%%%%%%%%%%%%%%%%%%%%%%%%%%%%%%%%%%%%%%%%%%%%%%%%%%%%%%%%%%%%%%%%%%%%%%%%%%%%%%
%%%%%%%%%%%%%%%%%%%%%%%%%%%%%%%%%%%%%%%%%%%%%%%%%%%%%%%%%%%%%%%%%%%%%%%%%%%%%%%%%
\begin{table}[h!]
\begin{center}
 \begin{tabular}{||c||l||}
 \hline
  Knot & \multicolumn{1}{|c||}{$Z_n(p)$} \\
  \hline  
$0_1$ &\fns $6\,{{\rm e}^{-8\,p}}+6\,{{\rm e}^{-4\,p}}$ \\
$3_1^+$ &\fns $24\,{{\rm e}^{-125\,p}}+480\,{{\rm e}^{-100\,p}}+696\,{{\rm e}^{-80\,p}}
+384\,{{\rm e}^{-75\,p}}$ \\
$4_1$   &\fns $12\,{{\rm e}^{-80\,p}}$ \\
$5_1^+$ &\fns $48\,{{\rm e}^{-210\,p}}+624\,{{\rm e}^{-180\,p}}+792\,{{\rm e}^{-175\,
p}}+432\,{{\rm e}^{-168\,p}}+4896\,{{\rm e}^{-150\,p}}+1776\,{{\rm e}^
{-144\,p}}$ \\
        &\fns $+2736\,{{\rm e}^{-140\,p}}+264\,{{\rm e}^{-125\,p}}+1536\,{{\rm e}^{-120\,p}}+504\,{{\rm e}^{-112\,p}}+456\,{{\rm e}^{-108\,p}}+
768\,{{\rm e}^{-105\,p}}$ \\
$5_2^+$ &\fns $144\,{{\rm e}^{-180\,p}}+144\,{{\rm e}^{-175\,p}}+2064\,{{\rm e}^{-150
\,p}}+48\,{{\rm e}^{-144\,p}}+912\,{{\rm e}^{-140\,p}}+24\,{{\rm e}^{-
125\,p}}+1056\,{{\rm e}^{-120\,p}}$ \\
        &\fns $+480\,{{\rm e}^{-112\,p}}$ \\
$6_1^+$ &\fns $48\,{{\rm e}^{-150\,p}}+24\,{{\rm e}^{-112\,p}}$ \\
$6_2^+$ &\fns $336\,{{\rm e}^{-216\,p}}+3600\,{{\rm e}^{-180\,p}}+576\,{{\rm e}^{-175
\,p}}+3168\,{{\rm e}^{-150\,p}}+432\,{{\rm e}^{-144\,p}}+144\,{{\rm e}
^{-125\,p}}$ \\
$6_3$   &\fns $48\,{{\rm e}^{-210\,p}}+528\,{{\rm e}^{-180\,p}}+240\,{{\rm e}^{-168\,
p}}+1968\,{{\rm e}^{-150\,p}}+336\,{{\rm e}^{-144\,p}}+144\,{{\rm e}^{
-125\,p}}+48\,{{\rm e}^{-120\,p}}$ \\
$7_1^+$ &\fns $24\,{{\rm e}^{-252\,p}}+144\,{{\rm e}^{-216\,p}}+144\,{{\rm e}^{-210\,
p}}+936\,{{\rm e}^{-180\,p}}+216\,{{\rm e}^{-175\,p}}$ \\
$7_2^+$ &\fns $24\,{{\rm e}^{-180\,p}}$ \\
$7_3^+$ &\fns $48\,{{\rm e}^{-294\,p}}+480\,{{\rm e}^{-252\,p}}+144\,{{\rm e}^{-245\,
p}}+1632\,{{\rm e}^{-240\,p}}+144\,{{\rm e}^{-225\,p}}+720\,{{\rm e}^{
-216\,p}}$ \\
        &\fns $+9312\,{{\rm e}^{-210\,p}}+4128\,{{\rm e}^{-200\,p}}+3000\,{{\rm e}^{-180\,p}}+1152\,{{\rm e}^{-175\,p}}+384\,{{\rm e}^{-168\,p}}+
816\,{{\rm e}^{-160\,p}}$ \\
        &\fns $+48\,{{\rm e}^{-150\,p}}+480\,{{\rm e}^{-144\,
p}}$ \\
$7_4^+$ &\fns $24\,{{\rm e}^{-343\,p}}+480\,{{\rm e}^{-294\,p}}+2016\,{{\rm e}^{-252
\,p}}+1008\,{{\rm e}^{-245\,p}}+912\,{{\rm e}^{-216\,p}}+2592\,{
{\rm e}^{-210\,p}}$ \\
        &\fns $+144\,{{\rm e}^{-196\,p}}+48\,{{\rm e}^{-192\,p}}+2544\,{{\rm e}^{-180\,p}}+1008\,{{\rm e}^{-175\,p}}+48\,{{\rm e}^{-168
\,p}}+384\,{{\rm e}^{-150\,p}}$ \\
$7_5^+$ &\fns $48\,{{\rm e}^{-252\,p}}+192\,{{\rm e}^{-240\,p}}+144\,{{\rm e}^{-225\,
p}}+144\,{{\rm e}^{-216\,p}}+1872\,{{\rm e}^{-210\,p}}+1632\,{{\rm e}^
{-200\,p}}$ \\
        &\fns $+2688\,{{\rm e}^{-180\,p}}+528\,{{\rm e}^{-175\,p}}+192\,{{\rm e}^{-168\,p}}+816\,{{\rm e}^{-160\,p}}+48\,{{\rm e}^{-150\,p}}+
480\,{{\rm e}^{-144\,p}}$ \\
$7_6^+$ &\fns $48\,{{\rm e}^{-150\,p}}$ \\
$7_7^+$ &\fns $24\,{{\rm e}^{-144\,p}}$ \\
\hline
 \end{tabular}
\end{center}
 \caption{Partition Functions: Minimal BCC lattice knots pressurized in a rectangular box.}
  \label{BCC-PartKnot-Box}
\end{table}
%%%%%%%%%%%%%%%%%%%%%%%%%%%%%%%%%%%%%%%%%%%%%%%%%%%%%%%%%%%%%%%%%%%%%%%%%%%%%%%%%
%%%%%%%%%%%%%%%%%%%%%%%%%%%%%%%%%%%%%%%%%%%%%%%%%%%%%%%%%%%%%%%%%%%%%%%%%%%%%%%%%

\subsubsection{Compressibility of $3_1^+$ in the FCC and BCC lattice:}  
Considering first the cases with the rectangular box volume $V_b$, the
partition functions of minimal length lattice trefoils in the FCC and BCC lattices
evaluate to 
\begin{eqnarray*}
\C{Z}_{FCC} (3_1^+) &=& 16{e^{-27p}}+48{e^{-24p}};\\
\C{Z}_{BCC} (3_1^+) &=& 24\,{{\rm e}^{-125\,p}}+480\,{{\rm e}^{-100\,p}}
+696\,{{\rm e}^{-80\,p}}+384\,{{\rm e}^{-75\,p}} .
\end{eqnarray*}

The expected volume of the lattice minimal knot at pressure $p$ is
\[ \LA V_b \RA
= 3\LH{\frac {{{\rm e}^{-51\,p}}}{ \left( 3\,{{\rm e}^{-27\,p}}+8\,{
{\rm e}^{-24\,p}} \right)  \left( {{\rm e}^{-27\,p}}+3\,{{\rm e}^{-24
\,p}} \right) }}\RH
\]
in the FCC lattice, and
\[ \LA V_b \RA
=  5\LH{\frac {25\,{{\rm e}^{-125\,p}}+400\,{{\rm e}^{-100\,p}}+464\,{
{\rm e}^{-80\,p}}+240\,{{\rm e}^{-75\,p}}}{{{\rm e}^{-125\,p}}+20\,{
{\rm e}^{-100\,p}}+29\,{{\rm e}^{-80\,p}}+16\,{{\rm e}^{-75\,p}}}}\RH
\]
in the BCC lattice.

The compressibilities in these cases are given by
\[ \beta = 
3\LH {\frac {{{\rm e}^{-51\,p}}}{ \left( 3\,{{\rm e}^{-27\,p}}+8\,{
{\rm e}^{-24\,p}} \right)  \left( {{\rm e}^{-27\,p}}+3\,{{\rm e}^{-24
\,p}} \right) }}\RH
 \]
in the FCC lattice, and 
\[ \fl \beta = 
5\LH{\frac {\hbox{\footnotesize
$500\,{{\rm e}^{-225\,p}}+2349\,{{\rm e}^{-205\,p}}+1600\,{
{\rm e}^{-200\,p}}+9280\,{{\rm e}^{-180\,p}}+8000\,{{\rm e}^{-175\,p}}
+464\,{{\rm e}^{-155\,p}}$}}{
\hbox{\footnotesize $\left( 25\,{{\rm e}^{-125\,p}}+400\,{
{\rm e}^{-100\,p}}+464\,{{\rm e}^{-80\,p}}+240\,{{\rm e}^{-75\,p}}
 \right)  A$}}}\RH
 \]
in the BCC lattice, where $A =  \left( {{\rm e}^{-125\,p}}+20\,{{\rm e}^{-100\,p}}+29\,{
{\rm e}^{-80\,p}}+16\,{{\rm e}^{-75\,p}} \right)$.

\begin{figure}[h!]
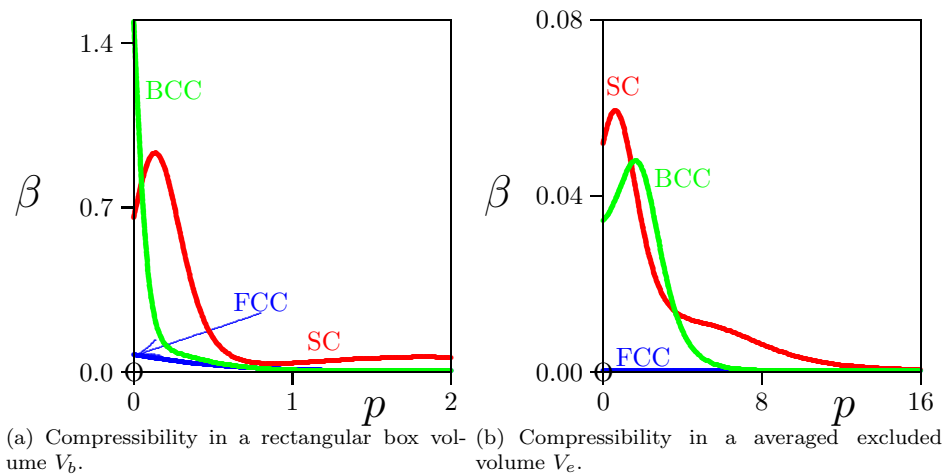

\centering
\subfigure[Compressibility in a rectangular box volume $V_b$.]{
\input{figure8A}
\label{FIG8A}}
\subfigure[Compressibility in a averaged excluded volume $V_e$.]{
\input{figure8B}
\label{FIG8B}}
\caption{Compressibility of the minimal lattice knots of type $3_1^+$
in the FCC and BCC lattices.  The compressibility in the SC lattice (see figures
\ref{FIG2} and \ref{FIG3}) are included for comparison.}
\label{FIG8} %ZXZ[FIG8]
\end{figure}

Plotting these compressibilities in the FCC and BCC lattices for $p\in [0,2]$ 
gives the curves in figure \ref{FIG8}.
%%%%%%%%%%%%%%%%%%%%%%%%%%%%%%%%%%%%%%%%%%%%%%%%
%%%%%%%%%%%%%%%%%%%%%%%%%%%%%%%%%%%%%%%%%%%%%%%%%%%%%%
If the average excluded volume $V_e$ is used instead, then the partition 
functions of minimal length lattice trefoils in the FCC and BCC lattices are 
\begin{eqnarray}
\C{Z}_{FCC} (3_1^+) &=& 64\,{{\rm e}^{-13/2\,p}};\\
\C{Z}_{BCC} (3_1^+) &=& 168\,{{\rm e}^{-16\,p}}+240\,{{\rm e}^{-{\frac {52}{3}}\,p}}+432\,{
{\rm e}^{-{\frac {50}{3}}\,p}}+144\,{{\rm e}^{-{\frac {53}{3}}\,p}} 
\nonumber \\
& &+288\,{{\rm e}^{-17\,p}}+72\,{{\rm e}^{-18\,p}}+96\,{{\rm e}^{-{\frac {
55}{3}}\,p}}+24\,{{\rm e}^{-{\frac {56}{3}}\,p}} \nonumber \\
& & +96\,{{\rm e}^{-{\frac {47}{3}}\,p}}+24\,{{\rm e}^{-{\frac {44}{3}}\,p}} .
\end{eqnarray}

The expected volume of the minimal lattice trefoil at pressure $p$ is equal
to $\LA V_e \RA = 13/2$ (independent on $p$) in the FCC lattice, and is 
a lengthy expression (dependent on $p$) in the BCC lattice.  The compressibility 
in the FCC lattice is identically zero, and in the BCC lattice is again a lengthy 
expression. Plotting these compressibilities in the FCC and BCC lattices 
for $p\in [0,2]$ gives the curves in  figure \ref{FIG8B}.

Observe that for both choices of the volumes $V_b$ and $V_e$, that 
$3_1^+$ is more compressible in the BCC lattice than in the FCC lattice.  This 
observation shows that the knot is much more tightly embedded in the 
FCC lattice, and that increasing pressure compress the knot less in the FCC lattice.

The global maxima of $\beta$ in figure \ref{FIG8A} (this is for the choice 
$V_b$) occurs at $p=0$ in the FCC and BCC lattices.  In this particular case, 
$\beta=1.17690\ldots$ in the FCC lattice and $\beta=1.48918\ldots$
in the BCC lattice at zero pressure.  The compressibility decreases monotonically 
with increasing $p>0$ in both cases.  If the volume $V_e$ is used instead, 
then $\beta=0$ for $p\geq 0$ in the FCC lattice (the lattice knot is 
incompressible), but there is a global maximum in $\beta$ at $p=1.62452\ldots$ 
where $\beta=0.04773\ldots$.

The maximum amount of useful work that can be performed by
letting a lattice knot type $3_1^+$ expand from its maximal
compressed state to its equilibrium at zero pressure in the FCC and
BCC lattices can be computed as before.  In these lattices,
$\C{W}_{3_1^+} = \log (4/3) = 0.28768\ldots$ in the FCC lattice,
and $\C{W}_{3_1^+} = \log (33/8) = 1.41707\ldots$ in the BCC
lattice.  If $V_e$ is used instead, then $\C{W}_{3_1^+} = 0$ in the 
FCC lattice, and $\C{W}_{3_1^+} = \log 66 = 4.18965\ldots$ in the BCC
lattice.

\subsubsection{The compressibility of minimal lattice knots in the FCC and BCC lattices:}  
\begin{figure}[h!]
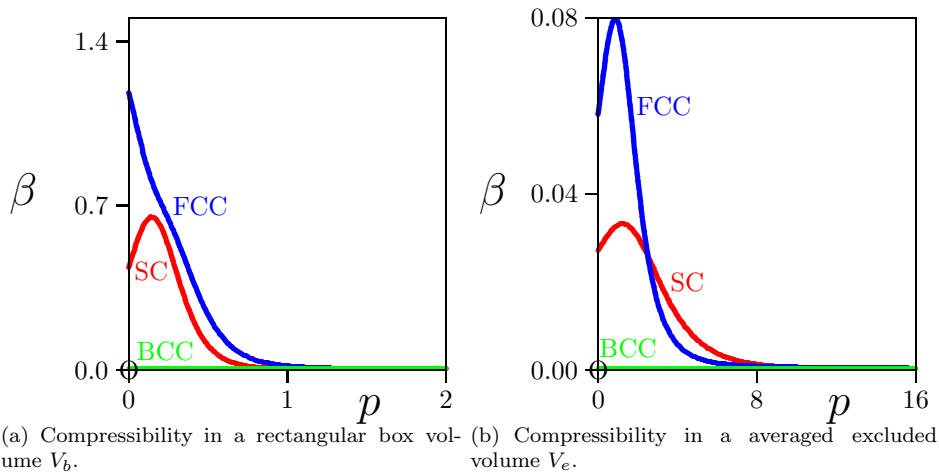

\centering
\subfigure[Compressibility in a rectangular box volume $V_b$.]{
\input{figure10A}
\label{FIG10A}}
\subfigure[Compressibility in a averaged excluded volume $V_e$.]{
\input{figure10B}
\label{FIG10B}}
\caption{Compressibility of the minimal lattice knots of type $4_1$
in the FCC and BCC lattices.  The compressibilities in the SC lattice 
(see figure \ref{FIG4}) are included for comparison.}
\label{FIG10} %%ZXZ[FIG10]
\end{figure}

The compressibilities of the figure eight knot $4_1$ are
illustrated in the FCC and BCC lattices for the choices of the volumes
$V_b$ and $V_e$ in figures \ref{FIG10}.
%%%%%%%%%%%%%%%%%%%%%%%%%%%%%%%%%%%%%%%%%%%%%%%%
%%%%%%%%%%%%%%%%%%%%%%%%%%%%%%%%%%%%%%%%%%%%%%%%%%%%%%
For the choice of volume $V_b$ in the FCC lattice, the maximum
compressibility of $4_1$ is at $p=0$, when $\beta=1.17690\ldots$,
but for $V_e$ one obtains a maximum at $p=0.86318\ldots$
where $\beta=0.07938\ldots$.  The knot is incompressible for
both choices $V_b$ and $V_e$ in the BCC lattice.

The maximum amount of work is $\C{W}_{4_1} = \log (233/16) =
2.67844\ldots$ in the FCC lattice for $V_b$, and $\C{W}_{4_1}
= \log (233/8) = 3.37159\ldots$ in the FCC lattice with the choice $V_e$.

Data on the compressibility of lattice knots in the FCC lattice
with the choice of volume $V_b$ is given in table \ref{CompKnot-Box-FCC}
and with the choice of volume $V_e$ in table \ref{CompKnot-Ve-FCC}.
%%%%%%%%%%%%%%%%%%%%%%%%%%%%%%%%%%%%%%%%%%%%%%%%%%%%%%%%%%%%%%%%%%%%%%%%%%%%%%%
%%%%%%%%%%%%%%%%%%%%%%%%%%%%%%%%%%%%%%%%%%%%%%%%%%%%%%%%%%%%%%%%%%%%%%%%%%%%%%%
\begin{table}[h!]
\begin{center}
 \begin{tabular}{||c||c||c|c||c|c||r||r||}
 \hline
  Knot & $\beta(0)$& $p_m$  & $\max\beta(p)$ & loc $p_m$ & loc $\max\beta(p)$ & $\C{W}_K$ & Min $V_b$  \\
  \hline  
$0_1$   &\fns $0$ &\fns $-$ &\fns $-$ &\fns $-$ &\fns $-$ &\fns $0$ &\fns $0$ \\
$3_1^+$ &\fns $0.06818$ &\fns $0$       &\fns $0.06818$ &\fns $-$ &\fns $-$  &\fns $0.28768$ &\fns $24$ \\
$4_1$   &\fns $1.17690$ &\fns $0$       &\fns $1.17690$ &\fns $-$ &\fns $-$  &\fns $2.67845$ &\fns $27$ \\
$5_1^+$ &\fns $0.54661$ &\fns $0.10381$ &\fns $0.72940$ &\fns $-$ &\fns $-$  &\fns $1.38629$ &\fns $36$ \\
$5_2^+$ &\fns $1.84036$ &\fns $0$       &\fns $1.84036$ &\fns $-$ &\fns $-$  &\fns $2.77259$ &\fns $40$ \\
$6_1^+$ &\fns $2.78110$ &\fns $0$       &\fns $2.78110$ &\fns $-$ &\fns $-$  &\fns $5.98141$ &\fns $40$ \\
$6_2^+$ &\fns $2.28512$ &\fns $0$       &\fns $2.28512$ &\fns $-$ &\fns $-$  &\fns $2.57452$ &\fns $45$ \\
$6_3$   &\fns $2.31350$ &\fns $0$       &\fns $2.31350$ &\fns $-$ &\fns $-$  &\fns $4.77819$ &\fns $36$ \\
$7_1^+$ &\fns $1.63106$ &\fns $0.02343$ &\fns $1.65022$ &\fns $0.46211$ &\fns $0.17066$  &\fns $4.44265$ &\fns $54$ \\
$7_2^+$ &\fns $2.63144$ &\fns $0$       &\fns $2.63144$ &\fns $-$ &\fns $-$  &\fns $3.76120$ &\fns $45$ \\
$7_3^+$ &\fns $2.10233$ &\fns $0$       &\fns $2.10233$ &\fns $-$ &\fns $-$  &\fns $1.60944$ &\fns $60$ \\
$7_4^+$ &\fns $0.03030$ &\fns $0.27976$ &\fns $0.04082$ &\fns $-$ &\fns $-$  &\fns $1.38629$ &\fns $96$ \\
$7_5^+$ &\fns $2.86502$ &\fns $0$       &\fns $2.86502$ &\fns $-$ &\fns $-$  &\fns $4.15191$ &\fns $54$ \\
$7_6^+$ &\fns $2.77693$ &\fns $0$       &\fns $2.77693$ &\fns $-$ &\fns $-$  &\fns $2.14007$ &\fns $48$ \\
$7_7^+$ &\fns $2.79374$ &\fns $0$       &\fns $2.79374$ &\fns $-$ &\fns $-$  &\fns $1.90954$ &\fns $48$ \\
$8_1^+$ &\fns $3.54811$ &\fns $0$       &\fns $3.54811$ &\fns $-$ &\fns $-$  &\fns $8.44779$ &\fns $48$ \\
$8_2^+$ &\fns $3.08699$ &\fns $0$       &\fns $3.08699$ &\fns $-$ &\fns $-$  &\fns $5.59306$ &\fns $54$ \\
$8_3$   &\fns $3.39870$ &\fns $0$       &\fns $3.39870$ &\fns $-$ &\fns $-$  &\fns $3.10109$ &\fns $60$ \\
$8_4^+$ &\fns $3.35418$ &\fns $0$       &\fns $3.35418$ &\fns $-$ &\fns $-$  &\fns $3.75654$ &\fns $60$ \\
$8_5^+$ &\fns $1.21929$ &\fns $0$       &\fns $1.21929$ &\fns $-$ &\fns $-$  &\fns $3.43399$ &\fns $63$ \\
$8_6^+$ &\fns $3.92441$ &\fns $0$       &\fns $3.92441$ &\fns $-$ &\fns $-$  &\fns $3.08089$ &\fns $60$ \\
$8_7^+$ &\fns $1.61860$ &\fns $0$       &\fns $1.61860$ &\fns $-$ &\fns $-$  &\fns $1.46634$ &\fns $64$ \\
$8_8^+$ &\fns $2.82549$ &\fns $0$       &\fns $2.82549$ &\fns $-$ &\fns $-$  &\fns $2.06369$ &\fns $60$ \\
$8_9$   &\fns $3.13220$ &\fns $0$       &\fns $3.13220$ &\fns $-$ &\fns $-$  &\fns $2.60269$ &\fns $60$ \\
$8_{10}^+$ &\fns $2.03727$ &\fns $0$       &\fns $2.03727$ &\fns $-$ &\fns $-$  &\fns $0.81093$ &\fns $80$ \\
$8_{11}^+$ &\fns $3.10730$ &\fns $0$       &\fns $3.10730$ &\fns $0.58372$ &\fns $0.18936$  &\fns $8.04093$ &\fns $54$ \\
$8_{12}$   &\fns $3.14959$ &\fns $0.00838$ &\fns $3.16225$ &\fns $-$ &\fns $-$  &\fns $2.65676$ &\fns $60$ \\
$8_{13}^+$ &\fns $3.25299$ &\fns $0$       &\fns $3.25299$ &\fns $-$ &\fns $-$  &\fns $5.24965$ &\fns $48$ \\
$8_{14}^+$ &\fns $3.07391$ &\fns $0$       &\fns $3.07391$ &\fns $-$ &\fns $-$  &\fns $2.12613$ &\fns $60$ \\
$8_{15}^+$ &\fns $2.67195$ &\fns $0$       &\fns $2.67195$ &\fns $-$ &\fns $-$  &\fns $4.84024$ &\fns $60$ \\
$8_{16}^+$ &\fns $1.88507$ &\fns $0$       &\fns $1.88507$ &\fns $-$ &\fns $-$  &\fns $1.21640$ &\fns $64$ \\
$8_{17}$   &\fns $1.81676$ &\fns $0$       &\fns $1.81676$ &\fns $-$ &\fns $-$  &\fns $1.21640$ &\fns $64$ \\
$8_{18}$   &\fns $3.11696$ &\fns $0.01503$ &\fns $3.16790$ &\fns $-$ &\fns $-$  &\fns $3.28916$ &\fns $45$ \\
$8_{19}^+$ &\fns $0.25869$ &\fns $0.17760$ &\fns $1.50198$ &\fns $-$ &\fns $-$  &\fns $3.13549$ &\fns $45$ \\
$8_{20}^+$ &\fns $1.73255$ &\fns $0$       &\fns $1.73255$ &\fns $-$ &\fns $-$  &\fns $5.95551$ &\fns $45$ \\
$8_{21}^+$ &\fns $2.06897$ &\fns $0$       &\fns $2.06897$ &\fns $-$ &\fns $-$  &\fns $2.39253$ &\fns $60$ \\
$9_1^+$    &\fns $2.00000$ &\fns $0.01014$ &\fns $2.02041$ &\fns $-$ &\fns $-$  &\fns $1.38629$ &\fns $80$ \\
$9_2^+$    &\fns $3.82335$ &\fns $0$       &\fns $3.82335$ &\fns $-$ &\fns $-$  &\fns $5.21577$ &\fns $60$ \\
$9_{42}^+$ &\fns $0$ &\fns $-$ &\fns $-$ &\fns $-$ &\fns $-$ &\fns $0$ &\fns $80$ \\
$9_{47}^+$ &\fns $0.92683$ &\fns $0.10758$ &\fns $1.04061$ &\fns $-$ &\fns $-$  &\fns $2.36712$ &\fns $64$ \\
$10_1^+$   &\fns $4.11636$ &\fns $0$       &\fns $4.11636$ &\fns $-$ &\fns $-$  &\fns $4.13116$ &\fns $80$ \\
$10_2^+$   &\fns $4.42190$ &\fns $0$       &\fns $4.42190$ &\fns $-$ &\fns $-$  &\fns $3.02852$ &\fns $80$ \\
  \hline
 \end{tabular}
\end{center}
 \caption{Compressibility of FCC Lattice Knots with volume $V_b$.}
  \label{CompKnot-Box-FCC} %ZXZ[CompKnot-Box-FCC]
\vspace{1mm}
\end{table}
%%%%%%%%%%%%%%%%%%%%%%%%%%%%%%%%%%%%%%%%%%%%%%%%%%%%%%%%%%%%%%%%%%%%%%%%%%%%%%%
%%%%%%%%%%%%%%%%%%%%%%%%%%%%%%%%%%%%%%%%%%%%%%%%%%%%%%%%%%%%%%%%%%%%%%%%%%%%%%%
%%%%%%%%%%%%%%%%%%%%%%%%%%%%%%%%%%%%%%%%%%%%%%%%%%%%%%%%%%%%%%%%%%%%%%%%%%%%%%%
%%%%%%%%%%%%%%%%%%%%%%%%%%%%%%%%%%%%%%%%%%%%%%%%%%%%%%%%%%%%%%%%%%%%%%%%%%%%%%%
\begin{table}[h!]
\begin{center}
 \begin{tabular}{||c||c||c|c||c|c||r||r||}
 \hline
 Knot & $\beta(0)$& $p_m$  & $\max\beta(p)$ & loc $p_m$ & loc $\max\beta(p)$ & $\C{W}_K$ & Min $V_e$  \\
  \hline  
$0_1$   &\fns $0$ &\fns $-$ &\fns $-$ &\fns $-$ &\fns $-$ &\fns $0$ &\fns $0$ \\
$3_1^+$ &\fns $0$ &\fns $-$ &\fns $-$ &\fns $-$ &\fns $-$ &\fns $0$ &\fns $13/2$ \\
$4_1$   &\fns $0.05773$ &\fns $0.86318$ &\fns $0.07939$ &\fns $-$ &\fns $-$ &\fns $3.37160$ &\fns $61/6$ \\
$5_1^+$ &\fns $0.01793$ &\fns $1.00305$ &\fns $0.02393$ &\fns $-$ &\fns $-$ &\fns $1.38629$ &\fns $37/3$ \\
$5_2^+$ &\fns $0.02808$ &\fns $0.39566$ &\fns $0.02899$ &\fns $-$ &\fns $-$ &\fns $2.77259$ &\fns $85/6$ \\
$6_1^+$ &\fns $0.05461$ &\fns $0.18826$ &\fns $0.05525$ &\fns $-$ &\fns $-$ &\fns $5.98141$ &\fns $101/6$ \\
$6_2^+$ &\fns $0.06128$ &\fns $0.57204$ &\fns $0.07799$ &\fns $-$ &\fns $-$ &\fns $4.65396$ &\fns $33/2$ \\
$6_3$   &\fns $0.03845$ &\fns $1.33926$ &\fns $0.07523$ &\fns $-$ &\fns $-$ &\fns $6.97541$ &\fns $107/6$ \\
$7_1^+$ &\fns $0.03763$ &\fns $1.23906$ &\fns $0.05985$ &\fns $-$ &\fns $-$ &\fns $4.44265$ &\fns $56/3$ \\
$7_2^+$ &\fns $0.04219$ &\fns $0.81100$ &\fns $0.05525$ &\fns $-$ &\fns $-$ &\fns $4.45435$ &\fns $41/2$ \\
$7_3^+$ &\fns $0.02067$ &\fns $0$       &\fns $0.02067$ &\fns $-$ &\fns $-$ &\fns $2.99573$ &\fns $22$ \\
$7_4^+$ &\fns $0$ &\fns $-$ &\fns $-$ &\fns $-$ &\fns $-$ &\fns $0$ &\fns $59/3$ \\
$7_5^+$ &\fns $0.04597$ &\fns $0.83134$ &\fns $0.06570$ &\fns $-$ &\fns $-$ &\fns $6.34914$ &\fns $127/6$ \\
$7_6^+$ &\fns $0.02878$ &\fns $0.60837$ &\fns $0.03174$ &\fns $-$ &\fns $-$ &\fns $4.62497$ &\fns $23$ \\
$7_7^+$ &\fns $0.00706$ &\fns $0$       &\fns $0.00706$ &\fns $-$ &\fns $-$ &\fns $2.60269$ &\fns $149/6$ \\
$8_1^+$ &\fns $0.07131$ &\fns $0.83333$ &\fns $0.08379$ &\fns $-$ &\fns $-$ &\fns $9.14094$ &\fns $68/3$ \\
$8_2^+$ &\fns $0.07810$ &\fns $0.70937$ &\fns $0.11184$ &\fns $-$ &\fns $-$ &\fns $7.09714$ &\fns $70/3$ \\
$8_3$   &\fns $0.05429$ &\fns $0.24897$ &\fns $0.05568$ &\fns $-$ &\fns $-$ &\fns $5.99146$ &\fns $51/2$ \\
$8_4^+$ &\fns $0.07784$ &\fns $0.64740$ &\fns $0.13305$ &\fns $-$ &\fns $-$ &\fns $6.05912$ &\fns $23$ \\
$8_5^+$ &\fns $0.03590$ &\fns $0.05611$ &\fns $0.03594$ &\fns $-$ &\fns $-$ &\fns $4.82028$ &\fns $27$ \\
$8_6^+$ &\fns $0.05098$ &\fns $0$       &\fns $0.05098$ &\fns $-$ &\fns $-$ &\fns $5.27811$ &\fns $157/6$ \\
$8_7^+$ &\fns $0.01714$ &\fns $1.00129$ &\fns $0.02788$ &\fns $-$ &\fns $-$ &\fns $3.25810$ &\fns $80/3$ \\
$8_8^+$ &\fns $0.02133$ &\fns $0$       &\fns $0.02133$ &\fns $-$ &\fns $-$ &\fns $4.14313$ &\fns $82/3$ \\
$8_9$   &\fns $0.03862$ &\fns $0.99643$ &\fns $0.05810$ &\fns $-$ &\fns $-$ &\fns $4.68213$ &\fns $76/3$ \\
$8_{10}^+$ &\fns $0.01719$ &\fns $0$       &\fns $0.01719$ &\fns $-$ &\fns $-$ &\fns $3.58352$ &\fns $55/2$ \\
$8_{11}^+$ &\fns $0.04565$ &\fns $0.90466$ &\fns $0.05686$ &\fns $-$ &\fns $-$ &\fns $8.73408$ &\fns $76/3$ \\
$8_{12}$   &\fns $0.04210$ &\fns $0.50800$ &\fns $0.04937$ &\fns $-$ &\fns $-$ &\fns $5.14166$ &\fns $82/3$ \\
$8_{13}^+$ &\fns $0.05485$ &\fns $0.93825$ &\fns $0.08329$ &\fns $-$ &\fns $-$ &\fns $8.65085$ &\fns $149/6$ \\
$8_{14}^+$ &\fns $0.02199$ &\fns $1.48654$ &\fns $0.02628$ &\fns $-$ &\fns $-$ &\fns $6.34564$ &\fns $167/6$ \\
$8_{15}^+$ &\fns $0.03477$ &\fns $0.53993$ &\fns $0.03777$ &\fns $-$ &\fns $-$ &\fns $6.63200$ &\fns $163/6$ \\
$8_{16}^+$ &\fns $0.02406$ &\fns $0.34559$ &\fns $0.02490$ &\fns $-$ &\fns $-$ &\fns $5.78074$ &\fns $85/3$ \\
$8_{17}$ &\fns $0.01195$ &\fns $0.63042$ &\fns $0.01246$ &\fns $-$ &\fns $-$ &\fns $4.68213$ &\fns $29$ \\
$8_{18}$ &\fns $0.04895$ &\fns $0$       &\fns $0.04895$ &\fns $0.85036$ &\fns $0.04679$ &\fns $6.06175$ &\fns $85/3$ \\
$8_{19}^+$ &\fns $0.01621$ &\fns $1.02391$ &\fns $0.01949$ &\fns $-$ &\fns $-$ &\fns $3.13549$ &\fns $56/3$ \\
$8_{20}^+$ &\fns $0.05023$ &\fns $0.98438$ &\fns $0.08007$ &\fns $-$ &\fns $-$ &\fns $6.24320$ &\fns $59/3$ \\
$8_{21}^+$ &\fns $0.02566$ &\fns $1.25537$ &\fns $0.03751$ &\fns $-$ &\fns $-$ &\fns $5.91889$ &\fns $133/6$ \\
$9_1^+$  &\fns $0.00758$ &\fns $0.03410$ &\fns $0.00758$ &\fns $-$ &\fns $-$ &\fns $1.38629$ &\fns $86/3$ \\
$9_2^+$  &\fns $0.07279$ &\fns $0.67316$ &\fns $0.08113$ &\fns $-$ &\fns $-$ &\fns $8.47387$ &\fns $161/6$ \\
$9_{42}^+$ &\fns $0.00028$ &\fns $0.02007$ &\fns $0.00028$ &\fns $-$ &\fns $-$ &\fns $0.69315$ &\fns $149/6$ \\
$9_{47}^+$ &\fns $0.01177$ &\fns $0.45773$ &\fns $0.01190$ &\fns $-$ &\fns $-$ &\fns $4.15888$ &\fns $89/3$ \\
$10_1^+$ &\fns $0.06685$ &\fns $0.55416$ &\fns $0.07921$ &\fns $-$ &\fns $-$ &\fns $7.38926$ &\fns $31$ \\
$10_2^+$ &\fns $0.06372$ &\fns $0.45145$ &\fns $0.08370$ &\fns $-$ &\fns $-$ &\fns $5.22575$ &\fns $32$ \\
 
  \hline
 \end{tabular}
\end{center}
 \caption{Compressibility of FCC Lattice Knots with volumes $V_e$.}
  \label{CompKnot-Ve-FCC} %ZXZ[CompKnot-Ve-FCC]
\vspace{1mm}
\end{table}
%%%%%%%%%%%%%%%%%%%%%%%%%%%%%%%%%%%%%%%%%%%%%%%%%%%%%%%%%%%%%%%%%%%%%%%%%%%%%%%
%%%%%%%%%%%%%%%%%%%%%%%%%%%%%%%%%%%%%%%%%%%%%%%%%%%%%%%%%%%%%%%%%%%%%%%%%%%%%%%
Observe that in most cases the maximal compressibility is at zero pressure, and that exceptions to this are sporadic in tables \ref{CompKnot-Box-FCC} and 
\ref{CompKnot-Ve-FCC}.  Even more sporadic are knot types with secondary local
maxima in the compressibility, as seen for example in figure \ref{FIG2} for
the trefoil knot -- in fact, with the choice of volume $V_b$ only
the knot types $7_1^+$ and $8_{11}^+$ have secondary local maximum in the
compressibility, and with $V_e$, only $8_{18}$.

Amongst knot types of $8$ crossing the knot types $8_{18}$, $8_{19}^+$ 
and $8_{20}^+$ can be realised in a rectangular box of volume $45$, 
small compared with the other eight crossing knots.  This is only slightly 
smaller than the minimum value of $V_b$ for $8_1^+$, which is $48$. 

The maximum work in table \ref{CompKnot-Box-FCC} is obtained for $8_1^+$, 
and $\C{W}_{8_1^+} = 8.44779 \ldots$.  Other knot  types which have high 
values for $\C{W}_{K}$ are $6_1^+$, $8_2^+$, $8_{11}^+$, 
$8_{13}^+$, $8_{20}^+$ and $9_2^+$.

If one considers the volume $V_e$ instead, then there is a steady increase
of $V_e$ with increasing knot complexity in table \ref{CompKnot-Ve-FCC}.
The maximum work in table \ref{CompKnot-Ve-FCC} is obtained for $8_1^+$,
and $\C{W}_{8_1^+} = 9.14094 \ldots$.  Other knot types which have high values
for $\C{W}_{K}$ are $6_3$, $8_2^+$, $8_{11}^+$, $8_{13}^+$, $9_2^+$ and 
$10_1^+$.

%%%%%%%%%%%%%%%%%%%%%%%%%%%%%%%%%%%%%%%%%%%%%%%%%%%%%%%%%%%%%%%%%%%%%%%%%%%%%%%
%%%%%%%%%%%%%%%%%%%%%%%%%%%%%%%%%%%%%%%%%%%%%%%%%%%%%%%%%%%%%%%%%%%%%%%%%%%%%%%
\begin{table}[h!]
\begin{center}
 \begin{tabular}{||c||c||c|c||c|c||r||r||}
 \hline
  Knot & $\beta(0)$& $p_m$  & $\max\beta(p)$ & loc $p_m$ & loc $\max\beta(p)$ & $\C{W}_K$ & Min $V_b$  \\
  \hline  
$0_1$   &\fns $0.66667$ &\fns $0.08664$ &\fns $0.68629$ &\fns $-$ &\fns $-$  &\fns $0.69315$ &\fns $4$ \\
$3_1^+$ &\fns $1.48918$ &\fns $0$       &\fns $1.48918$ &\fns $-$ &\fns $-$  &\fns $1.41707$ &\fns $75$ \\
$4_1$   &\fns $0$       &\fns $-$       &\fns $-$       &\fns $-$ &\fns $-$  &\fns $-$       &\fns $80$ \\
$5_1^+$ &\fns $2.58695$ &\fns $0.02790$ &\fns $2.78116$ &\fns $-$ &\fns $-$  &\fns $2.96075$ &\fns $105$ \\
$5_2^+$ &\fns $2.08587$ &\fns $0.00813$ &\fns $2.09679$ &\fns $-$ &\fns $-$  &\fns $2.31747$ &\fns $112$ \\
$6_1^+$ &\fns $2.33657$ &\fns $0.02208$ &\fns $2.77037$ &\fns $-$ &\fns $-$  &\fns $1.09861$ &\fns $112$ \\
$6_2^+$ &\fns $2.06977$ &\fns $0$       &\fns $2.06977$ &\fns $-$ &\fns $-$  &\fns $4.04888$ &\fns $125$ \\
$6_3$   &\fns $1.54825$ &\fns $0$       &\fns $1.54825$ &\fns $0.09268$ &\fns $1.34564$  &\fns $4.23411$ &\fns $120$ \\
$7_1^+$ &\fns $1.37260$ &\fns $0$       &\fns $1.37260$ &\fns $0.28854$ &\fns $0.03530$  &\fns $1.91365$ &\fns $175$ \\
$7_2^+$ &\fns $0$       &\fns $-$       &\fns $-$       &\fns $-$ &\fns $-$  &\fns $-$       &\fns $180$ \\
$7_3^+$ &\fns $2.41600$ &\fns $0.04325$ &\fns $3.10143$ &\fns $-$ &\fns $-$  &\fns $3.84695$ &\fns $144$ \\
$7_4^+$ &\fns $5.70413$ &\fns $0$       &\fns $5.70413$ &\fns $-$ &\fns $-$  &\fns $3.37374$ &\fns $150$ \\
$7_5^+$ &\fns $2.49925$ &\fns $0.01719$ &\fns $2.56136$ &\fns $-$ &\fns $-$  &\fns $2.90690$ &\fns $144$ \\
$7_6^+$ &\fns $0$       &\fns $-$       &\fns $-$       &\fns $-$ &\fns $-$  &\fns $-$       &\fns $150$ \\
$7_7^+$ &\fns $0$       &\fns $-$       &\fns $-$       &\fns $-$ &\fns $-$  &\fns $-$       &\fns $144$ \\
$8_1^+$ &\fns $2.25572$ &\fns $0.05046$ &\fns $5.70492$ &\fns $-$ &\fns $-$  &\fns $3.43399$ &\fns $144$ \\
$8_2^+$ &\fns $3.20299$ &\fns $0$       &\fns $3.20299$ &\fns $-$ &\fns $-$  &\fns $7.11477$ &\fns $168$ \\
$8_3$   &\fns $1.96612$ &\fns $0.03662$ &\fns $5.29029$ &\fns $-$ &\fns $-$  &\fns $2.19722$ &\fns $144$ \\
$8_4^+$ &\fns $5.43599$ &\fns $0$       &\fns $5.43599$ &\fns $0.38809$ &\fns $0.10046$  &\fns $7.57968$ &\fns $168$ \\
$8_5^+$ &\fns $3.14193$ &\fns $0$       &\fns $3.14193$ &\fns $-$ &\fns $-$  &\fns $2.01653$ &\fns $175$ \\
$8_6^+$ &\fns $3.44004$ &\fns $0$       &\fns $3.44004$ &\fns $-$ &\fns $-$  &\fns $5.23644$ &\fns $160$ \\
$8_7^+$ &\fns $3.37683$ &\fns $0$       &\fns $3.37683$ &\fns $0.10270$ &\fns $1.94863$  &\fns $6.90475$ &\fns $160$ \\
$8_8^+$ &\fns $3.49969$ &\fns $0$       &\fns $3.49969$ &\fns $-$ &\fns $-$  &\fns $4.79027$ &\fns $168$ \\
$8_9$   &\fns $2.59394$ &\fns $0$       &\fns $2.59394$ &\fns $0.05510$ &\fns $1.94102$  &\fns $3.16969$ &\fns $175$ \\
$8_{10}^+$ &\fns $2.24347$ &\fns $0$       &\fns $2.24347$ &\fns $-$ &\fns $-$  &\fns $2.55205$ &\fns $180$ \\
$8_{11}^+$ &\fns $2.41677$ &\fns $0$       &\fns $2.41677$ &\fns $-$ &\fns $-$  &\fns $5.80212$ &\fns $175$ \\
$8_{12}$   &\fns $0$       &\fns $-$       &\fns $-$       &\fns $-$ &\fns $-$  &\fns $-$       &\fns $144$ \\
$8_{13}^+$ &\fns $4.10106$ &\fns $0$       &\fns $4.10106$ &\fns $-$ &\fns $-$  &\fns $5.21221$ &\fns $168$ \\
$8_{14}^+$ &\fns $3.13157$ &\fns $0$       &\fns $3.13157$ &\fns $-$ &\fns $-$  &\fns $5.86647$ &\fns $160$ \\
$8_{15}^+$ &\fns $1.73039$ &\fns $0.05233$ &\fns $2.12601$ &\fns $-$ &\fns $-$  &\fns $3.39002$ &\fns $168$ \\
$8_{16}^+$ &\fns $0.97928$ &\fns $0$       &\fns $0.97928$ &\fns $-$ &\fns $-$  &\fns $1.29928$ &\fns $175$ \\
$8_{17}$   &\fns $0.92879$ &\fns $0$       &\fns $0.92879$ &\fns $-$ &\fns $-$  &\fns $0.30538$ &\fns $180$ \\
$8_{18}$   &\fns $4.70675$ &\fns $0$       &\fns $4.70675$ &\fns $0.16392$ &\fns $2.02337$  &\fns $6.59987$ &\fns $144$ \\
$8_{19}^+$ &\fns $1.69240$ &\fns $0.03863$ &\fns $2.08744$ &\fns $-$ &\fns $-$  &\fns $1.72385$ &\fns $144$ \\
$8_{20}^+$ &\fns $3.41192$ &\fns $0$       &\fns $3.41192$ &\fns $-$ &\fns $-$  &\fns $4.96634$ &\fns $150$ \\
$8_{21}^+$ &\fns $2.03630$ &\fns $0$       &\fns $2.03630$ &\fns $-$ &\fns $-$  &\fns $2.26868$ &\fns $175$ \\
$9_1^+$ &\fns $4.05666$ &\fns $0$         &\fns $4.05666$ &\fns $-$ &\fns $-$  &\fns $5.92604$ &\fns $180$ \\
$9_2^+$ &\fns $4.80856$ &\fns $0$         &\fns $4.80856$ &\fns $0.08777$ &\fns $2.46128$  &\fns $5.66296$ &\fns $168$ \\
$9_{42}^+$ &\fns $1.80198$ &\fns $0$         &\fns $1.80198$ &\fns $-$ &\fns $-$  &\fns $3.34990$ &\fns $175$ \\
$9_{47}^+$ &\fns $2.10389$ &\fns $0$         &\fns $2.10389$ &\fns $0.09644$ &\fns $1.64930$  &\fns $5.06189$ &\fns $175$ \\
$10_1^+$ &\fns $3.33858$ &\fns $0$       &\fns $3.33858$ &\fns $-$ &\fns $-$  &\fns $1.79176$ &\fns $210$ \\
$10_2^+$ &\fns $2.72186$ &\fns $0.01441$ &\fns $2.79899$ &\fns $-$ &\fns $-$  &\fns $2.14251$ &\fns $240$ \\

  \hline
 \end{tabular}
\end{center}
 \caption{Compressibility of BCC Lattice Knots with volumes $V_b$.}
  \label{CompKnot-Box-BCC} %ZXZ[CompKnot-Box-BCC]
\vspace{1mm}
\end{table}
%%%%%%%%%%%%%%%%%%%%%%%%%%%%%%%%%%%%%%%%%%%%%%%%%%%%%%%%%%%%%%%%%%%%%%%%%%%%%%%
%%%%%%%%%%%%%%%%%%%%%%%%%%%%%%%%%%%%%%%%%%%%%%%%%%%%%%%%%%%%%%%%%%%%%%%%%%%%%%%
%%%%%%%%%%%%%%%%%%%%%%%%%%%%%%%%%%%%%%%%%%%%%%%%%%%%%%%%%%%%%%%%%%%%%%%%%%%%%%%
%%%%%%%%%%%%%%%%%%%%%%%%%%%%%%%%%%%%%%%%%%%%%%%%%%%%%%%%%%%%%%%%%%%%%%%%%%%%%%%
\begin{table}[h!]
\begin{center}
 \begin{tabular}{||c||c||c|c||c|c||r||r||}
 \hline
 Knot & $\beta(0)$& $p_m$  & $\max\beta(p)$ & loc $p_m$ & loc $\max\beta(p)$ & $\C{W}_K$ & Min $V_e$ \\
  \hline  
$0_1$   &\fns $0$ &\fns $-$ &\fns $-$ &\fns $-$ &\fns $-$ &\fns $0$ &\fns $0$ \\
$3_1^+$ &\fns $0.03402$ &\fns $1.62452$ &\fns $0.04773$ &\fns $-$ &\fns $-$ &\fns $4.18965$ &\fns $44/3$ \\
$4_1$   &\fns $0$ &\fns $-$ &\fns $-$   &\fns $-$ &\fns $-$ &\fns $0$ &\fns $64/3$ \\
$5_1^+$ &\fns $0.06553$ &\fns $1.22854$ &\fns $0.10245$ &\fns $-$ &\fns $-$ &\fns $5.32788$ &\fns $80/3$ \\
$5_2^+$ &\fns $0.05674$ &\fns $0$       &\fns $0.05674$ &\fns $-$ &\fns $-$ &\fns $4.62006$ &\fns $91/3$ \\
$6_1^+$ &\fns $0.00629$ &\fns $0$       &\fns $0.00629$ &\fns $-$ &\fns $-$ &\fns $0.40547$ &\fns $35$ \\
$6_2^+$ &\fns $0.05241$ &\fns $0$       &\fns $0.05241$ &\fns $1.37261$ &\fns $0.03755$ &\fns $5.14749$ &\fns $116/3$ \\
$6_3$   &\fns $0.03892$ &\fns $0.79096$ &\fns $0.05619$ &\fns $-$ &\fns $-$ &\fns $4.23411$ &\fns $116/3$ \\
$7_1^+$ &\fns $0.03748$ &\fns $0$       &\fns $0.03748$ &\fns $-$ &\fns $-$ &\fns $1.71298$ &\fns $124/3$ \\
$7_2^+$ &\fns $0$ &\fns $-$ &\fns $-$   &\fns $-$ &\fns $-$ &\fns $0$ &\fns $44$ \\
$7_3^+$ &\fns $0.06347$ &\fns $0$       &\fns $0.06347$ &\fns $0.22071$ &\fns $0.06115$ &\fns $6.14954$ &\fns $134/3$ \\
$7_4^+$ &\fns $0.09845$ &\fns $0.57690$ &\fns $0.38184$ &\fns $-$ &\fns $-$ &\fns $5.45318$ &\fns $40$ \\
$7_5^+$ &\fns $0.06912$ &\fns $0$       &\fns $0.06912$ &\fns $-$ &\fns $-$ &\fns $5.20949$ &\fns $48$ \\
$7_6^+$ &\fns $0$ &\fns $-$ &\fns $-$   &\fns $-$ &\fns $-$ &\fns $0$ &\fns $47$ \\
$7_7^+$ &\fns $0$ &\fns $-$ &\fns $-$   &\fns $-$ &\fns $-$ &\fns $0$ &\fns $48$ \\
$8_1^+$ &\fns $0.09074$ &\fns $0.23262$ &\fns $0.11279$ &\fns $-$ &\fns $-$ &\fns $2.74084$ &\fns $49$ \\
$8_2^+$ &\fns $0.09247$ &\fns $0.71950$ &\fns $0.17808$ &\fns $-$ &\fns $-$ &\fns $7.80792$ &\fns $152/3$ \\
$8_3$   &\fns $0.00856$ &\fns $0$       &\fns $0.00856$ &\fns $-$ &\fns $-$ &\fns $0.58779$ &\fns $152/3$ \\
$8_4^+$ &\fns $0.13367$ &\fns $0.39542$ &\fns $0.26081$ &\fns $-$ &\fns $-$ &\fns $6.88653$ &\fns $148/3$ \\
$8_5^+$ &\fns $0.07155$ &\fns $0$       &\fns $0.07155$ &\fns $-$ &\fns $-$ &\fns $5.03695$ &\fns $167/3$ \\
$8_6^+$ &\fns $0.06902$ &\fns $0.65504$ &\fns $0.10783$ &\fns $-$ &\fns $-$ &\fns $5.23644$ &\fns $54$ \\
$8_7^+$ &\fns $0.08047$ &\fns $0.57910$ &\fns $0.13155$ &\fns $-$ &\fns $-$ &\fns $6.90475$ &\fns $52$ \\
$8_8^+$ &\fns $0.11540$ &\fns $0.46546$ &\fns $0.11558$ &\fns $-$ &\fns $-$ &\fns $6.58203$ &\fns $54$ \\
$8_9$   &\fns $0.06412$ &\fns $0$       &\fns $0.06412$ &\fns $-$ &\fns $-$ &\fns $4.77912$ &\fns $169/3$ \\
$8_{10}^+$ &\fns $0.04289$ &\fns $0.77248$ &\fns $0.05574$ &\fns $-$ &\fns $-$ &\fns $5.44242$ &\fns $166/3$ \\
$8_{11}^+$ &\fns $0.07035$ &\fns $0.93770$ &\fns $0.07418$ &\fns $-$ &\fns $-$ &\fns $5.80212$ &\fns $166/3$ \\
$8_{12}$   &\fns $0$ &\fns $-$ &\fns $-$   &\fns $-$ &\fns $-$ &\fns $0$ &\fns $176/3$ \\
$8_{13}^+$ &\fns $0.05706$ &\fns $0.71565$ &\fns $0.07935$ &\fns $-$ &\fns $-$ &\fns $5.90536$ &\fns $56$ \\
$8_{14}^+$ &\fns $0.09919$ &\fns $0$       &\fns $0.09919$ &\fns $-$ &\fns $-$ &\fns $5.86647$ &\fns $170/3$ \\
$8_{15}^+$ &\fns $0.05556$ &\fns $0.86305$ &\fns $0.06846$ &\fns $-$ &\fns $-$ &\fns $4.48864$ &\fns $57$ \\
$8_{16}^+$ &\fns $0.08002$ &\fns $0.19909$ &\fns $0.08208$ &\fns $-$ &\fns $-$ &\fns $3.09104$ &\fns $58$ \\
$8_{17}$   &\fns $0.04271$ &\fns $0$       &\fns $0.04271$ &\fns $-$ &\fns $-$ &\fns $2.94444$ &\fns $58$ \\
$8_{18}$   &\fns $0.16917$ &\fns $0$       &\fns $0.16917$ &\fns $-$ &\fns $-$ &\fns $5.21358$ &\fns $191/3$ \\
$8_{19}^+$ &\fns $0.17811$ &\fns $0$       &\fns $0.17811$ &\fns $-$ &\fns $-$ &\fns $2.44777$ &\fns $42$ \\
$8_{20}^+$ &\fns $0.11282$ &\fns $0.07141$ &\fns $0.11314$ &\fns $-$ &\fns $-$ &\fns $7.79955$ &\fns $134/3$ \\
$8_{21}^+$ &\fns $0.04977$ &\fns $0$       &\fns $0.04977$ &\fns $-$ &\fns $-$ &\fns $2.67415$ &\fns $148/3$ \\
$9_1^+$    &\fns $0.07219$ &\fns $0$       &\fns $0.07219$ &\fns $-$ &\fns $-$ &\fns $7.43011$ &\fns $57$ \\
$9_2^+$    &\fns $0.08763$ &\fns $0$       &\fns $0.08763$ &\fns $0.89024$ &\fns $0.08575$ &\fns $5.66296$ &\fns $170/3$ \\
$9_{42}^+$ &\fns $0.04013$ &\fns $0.46670$ &\fns $0.05049$ &\fns $-$ &\fns $-$ &\fns $4.04305$ &\fns $157/3$ \\
$9_{47}^+$ &\fns $0.09375$ &\fns $0$       &\fns $0.09375$ &\fns $-$ &\fns $-$ &\fns $6.56597$ &\fns $63$ \\
$10_1^+$   &\fns $0.05363$ &\fns $0.23863$ &\fns $0.06286$ &\fns $-$ &\fns $-$ &\fns $1.79176$ &\fns $64$ \\
$10_2^+$   &\fns $0.06635$ &\fns $0.89586$ &\fns $0.09463$ &\fns $-$ &\fns $-$ &\fns $6.01372$ &\fns $200/3$ \\
  \hline
 \end{tabular}
\end{center}
 \caption{Compressibility of BCC Lattice Knots with volumes $V_e$.}
  \label{CompKnot-Ve-BCC} %ZXZ[CompKnot-Ve-BCC]
\vspace{1mm}
\end{table}
%%%%%%%%%%%%%%%%%%%%%%%%%%%%%%%%%%%%%%%%%%%%%%%%%%%%%%%%%%%%%%%%%%%%%%%%%%%%%%%
%%%%%%%%%%%%%%%%%%%%%%%%%%%%%%%%%%%%%%%%%%%%%%%%%%%%%%%%%%%%%%%%%%%%%%%%%%%%%%%

Data on the compressibility of lattice knots in the BCC lattice
with the choice of volume $V_b$ are given in table \ref{CompKnot-Box-BCC}
and with the choice of volume $V_e$ in table \ref{CompKnot-Ve-BCC}.

In most cases the maximal compressibility is at zero pressure, but there
are some exceptions to this in tables \ref{CompKnot-Box-BCC} and 
\ref{CompKnot-Ve-BCC}.  For example, the maximum compressibility
of the unknot is at $p=0.08664\ldots$ for the choice $V_b$ in table
\ref{CompKnot-Box-BCC}.  

There are also some knot types with secondary local maxima in the 
compressibility, similar to the case seen in figure \ref{FIG2} for the
trefoil in the SC lattice with the choice of $V_b$ as volume.
With the choice $V_b$ in the BCC, these include  $6_3$, $7_1^+$, 
$8_4^+$, $8_7^+$, $8_9$, $8_{18}$, $9_2^+$ and $9_{47}^+$, and 
for the choice $V_e$, $6_2^+$, $7_3^+$ and $9_2^+$.

Amongst knot types of eight crossings the knot types $8_{1}^+$, $8_{3}$, 
$8_{12}$, $8_{18}$ and $8_{19}^+$ can be realised in a rectangular box 
of volume $144$; this is smaller than other eight crossing knots, and
these knot types have compact minimal length conformations in the BCC.
As one would expect, with some exceptions, these knot types are also 
less compressible than eight crossing knots.

The maximum work in table \ref{CompKnot-Box-BCC} is obtained for $8_4^+$, 
and $\C{W}_{8_4^+} = 7.57968\ldots$.  Other knot  types which have high 
values for $\C{W}_{K}$ are $8_2^+$, $8_7^+$ and $8_{18}$.

If one considers the volume $V_e$ instead, then there is a steady increase
of $V_e$ with increasing knot complexity in table \ref{CompKnot-Ve-BCC}.
Amongst knot types of eight crossing the knot types $8_{19}^+$ and $8_{20}^+$
can be realised with averaged excluded volume $V_e$ equal to $42$ and 
$134/3$, this is small when compared to other eight crossing knots. 

The maximum work in table \ref{CompKnot-Ve-BCC} is obtained for $8_2^+$, and 
$\C{W}_{8_2^+} = 7.80792 \ldots$.  Other knot types which has high values
for $\C{W}_{K}$ are $8_{20}^+$ and $9_1^+$.

%%%%%%%%%%%%%%%%%%%%%%%%%%%%%%%%%%%%%%%%%%%%%%%%%%%%%%%%%%%%%%%%%%%%%%%%%
%%%%%%%%%%%%%%%%%%%%%%%%%%%%%%%%%%%%%%%%%%%%%%%%%%%%%%%%%%%%%%%%%%%%%%%%%
\subsection{Discussion of Results}

The compressibility data of SC lattice knots in tables \ref{CompKnot-Box}
and \ref{CompKnot-Box-CH} indicate that different knot types may have
very different properties.  That is, the compressibilities are
functions of the topologies of the embedded polygons, as well as of
the geometry of the polygons, since the results are also dependent
on the lattice, as seen for example when comparing the data in tables
\ref{CompKnot-Box}, \ref{CompKnot-Box-FCC} and \ref{CompKnot-Box-BCC}.

Generally, the minimal length lattice knots were also more compressible
in the volume $V_b$, compared to the average excluded volume $V_e$
which is a more close-fitting envelope about the polygon.  This should
be due to the fact that there is extra space for the compressed
polygon to expand into near the corners of the rectangular box in the
volume $V_b$, and this is not available in the volume $V_e$.

\begin{figure}[t!]
\input{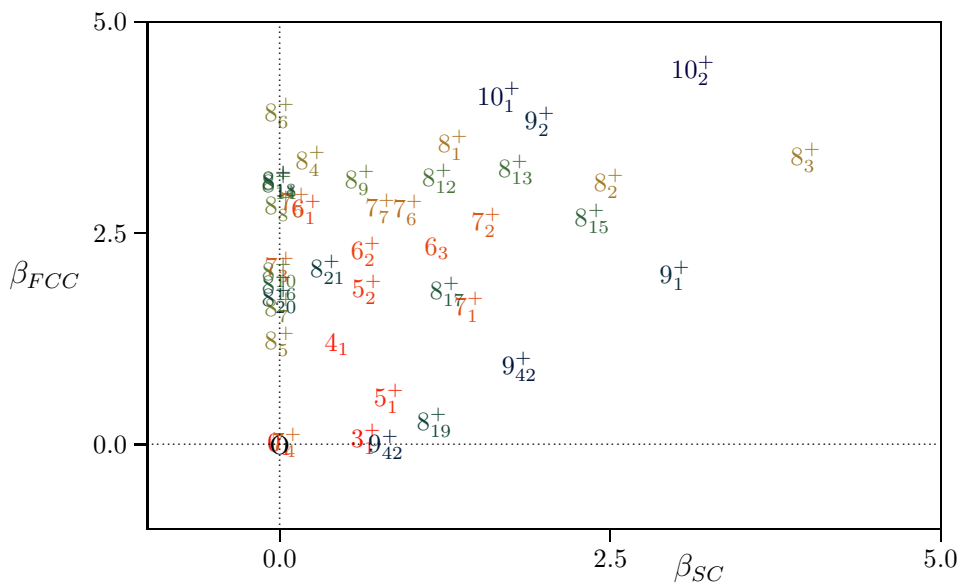}
\caption{Scatter plot of zero pressure compressibilities ($V_b)$ of minimal 
lattice knots in the SC and FCC lattices.  Generally, the lattice knots
are more compressible in the FCC, although there are exceptions to this, 
these would be the knot types which lay below the diagonal in this 
diagram.}
\label{ScatterBetaVbSCFCC} %ZXZ[ScatterBetaVbSCFCC]
\end{figure}

\begin{figure}[h!]
\input{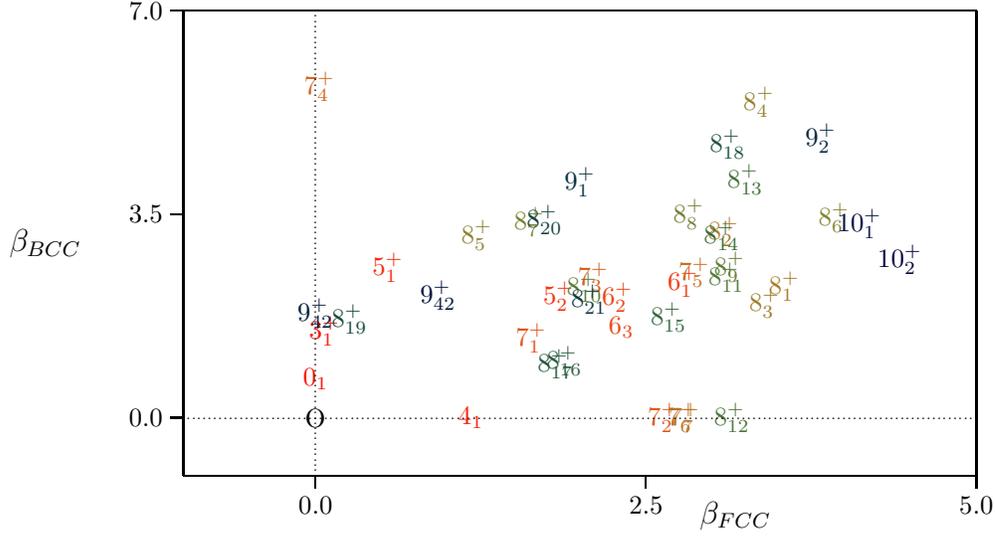}
\caption{Scatter plot of zero pressure compressibilities $(V_b)$ of minimal 
lattice knots in the FCC and BCC lattices.  Generally, the lattice knots
are more compressible in the FCC.}
\label{ScatterBetaVbFCCBCC} %ZXZ[ScatterBetaVbFCCBCC]
\end{figure}

There appears to the little correlation between the zero pressure
compressibility of minimal lattice knots in the three lattices.
For example, in figures \ref{ScatterBetaVbSCFCC} and 
\ref{ScatterBetaVbFCCBCC} two scatter plots of the zero 
pressure compressibility computed from the rectangular box volume
$V_b$ are given, showing a wide dispersion of points and little correlation 
between the data in different lattices.
%%%%%%%%%%%%%%%%%%%%%%%%%%%%%%%%%%%%%%%%%%%%%%%%%%
%%%%%%%%%%%%%%%%%%%%%%%%%%%%%%%%%%%%%%%%%%%%%%%%%%%
%%%%%%%%%%%%%%%%%%%%%%%%%%%%%%%%%%%%%%%%%%%%%%%%%%
%%%%%%%%%%%%%%%%%%%%%%%%%%%%%%%%%%%%%%%%%%%%%%%%%%%
There are some generic trends visible in these plots, for example,
compressibility tends to increase with crossing number in at least one
of the lattices (and knots with low crossing number are likely to be
closer to the origin in these plots).

\begin{figure}[t]
\input{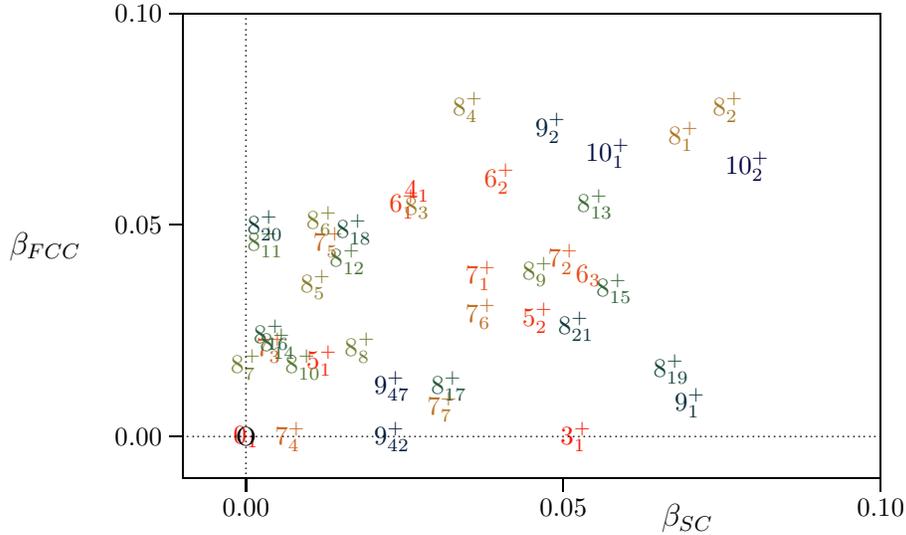}
\caption{Scatter plot of zero pressure compressibilities of minimal 
lattice knots computed from the volume $V_e$ in the FCC and BCC 
lattices.}
\label{ScatterBetaVeSCFCC} %ZXZ[ScatterBetaVeSCFCC]
\end{figure}

The situation is similar if the compressibilities are instead
computed by consider the more close fitting volume $V_e$.  In
figure \ref{ScatterBetaVeSCFCC} a scatter plot of the zero pressure
compressibilities in the SC and FCC lattices, computed from $V_e$,
are illustrated.  The wide dispersion of the points are similar
to the results seen for the volume $V_b$.
%%%%%%%%%%%%%%%%%%%%%%%%%%%%%%%%%%%%%%%%%%%%%%%%%%
%%%%%%%%%%%%%%%%%%%%%%%%%%%%%%%%%%%%%%%%%%%%%%%%%%%

\begin{figure}[h!]
\input{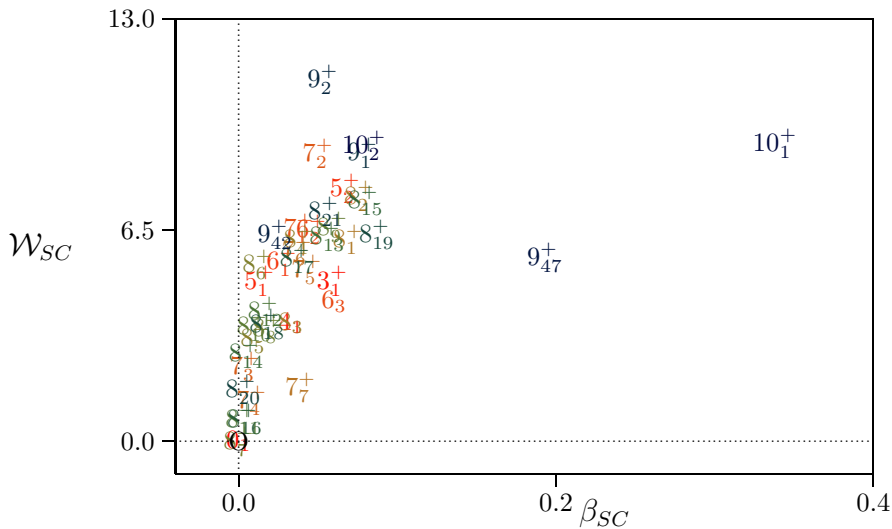}
\caption{Scatter plot of maximum compressibilities of minimal 
lattice knots and the total amount of work $\C{W}_L$
computed from the volume $V_e$ in the SC lattice.  The data show
a close correlation, with the knot types $9_{47}^+$ and $10_1^+$
clear outliers in this graph.}
\label{ScatterBetaWVeSC} %ZXZ[ScatterBetaWVeSC]
\end{figure}

Examination of the data on $\C{W}_K$ in each of the lattices and
for the knot types, shows that there are some correlations
between compressibility and the total amount of work.  For example,
in figure \ref{ScatterBetaWVeSC} a strong correlation is observed
between the maximum compressibility $\beta_m$ and $\C{W}_K$ of knot types
in the SC lattice computed from the volume $V_e$.
%%%%%%%%%%%%%%%%%%%%%%%%%%%%%%%%%%%%%%%%%%%%%%%%%%%
%%%%%%%%%%%%%%%%%%%%%%%%%%%%%%%%%%%%%%%%%%%%%%%%%%%%
Similar graphs are obtained in the other lattices and also 
for the choice of $V_b$ as containing volume.  There are somewhat
weaker correlations between zero pressure compressibility and 
$\C{W}_K$ in all cases.

The data in tables \ref{CompKnot-Box}, \ref{CompKnot-Box-FCC},
\ref{CompKnot-Ve-FCC}, \ref{CompKnot-Box-BCC} and \ref{CompKnot-Ve-BCC}
show an association between knot types having secondary local
maxima in their compressibility curves, and having high total
value of $\C{W}_K$.  In table \ref{CompKnot-Box} the knot types
$3_1^+$, $5_1^+$, $6_2^+$, $9_2^+$ and $10_1^+$ have local maxima
in $\beta_{SC}$ (the case for $3_1^+$ is illustrated in figure \ref{FIG2}),
and all have higher values of $\C{W}_K$ than comparable knots. A similar
pattern can be seen in the other other tables.

%%%%%%%%%%%%%%%%%%%%%%%%%%%%%%%%%%%%%%%%%%%%%%%%%%%%%%%%%%%%%%%%%%%%%%%%%
%%%%%%%%%%%%%%%%%%%%%%%%%%%%%%%%%%%%%%%%%%%%%%%%%%%%%%%%%%%%%%%%%%%%%%%%%
\section{Conclusions}

In this paper the compressibility of minimal length lattice knots were
examined in three lattices, namely the SC, the FCC and the BCC lattices.
Generally, the compressibility properties were found to be dependent
on the lattice as well as the knot type.  Since the lattice impose
a geometric constraint on the lattice knots, the compressibility 
is determined to be a function of the geometry.

More generally, different knot types were also found to have very
different compressibility properties in each of the lattices.  This
is for example illustrated in figure \ref{FIG5} and \ref{FIG6} for
SC lattice knots;  the compressibilities of knot types of roughly 
the same geometric and topological complexity (for example, about the 
same minimal length and with the same crossover numbers) are seen
to have different compressibilities with increasing forces.

We examined the compressibilities by using two notions of a confining
space for the lattice knots, firstly the smallest rectangular box
volume $V_b$ containing the polygons, and secondly an average excluded
volume $V_e$ computed by slicing the knot in slabs, taking the convex hulls
of those slabs, and then their union. $V_e$ is a very 
tight envelope around the lattice knot.  

In figure \ref{VminSCSC}
we show that there is a strong correlation in knot type between $V_b$ and
$V_e$ in the SC lattice, and the convex hull volume is thus similarly
highly correlated with both $V_b$ and $V_e$.  Similar results were obtained
in the FCC and BCC lattices.  Thus, for many knot types the
compressibilities $\beta$ were similarly behaved for either choice of the
volume, as one may see in figures \ref{FIG2} and \ref{FIG3} for the
knot type $3_1$ and in figure \ref{FIG4} for the knot type $4_1$
in the SC lattice.  This was not universally the case however, as for
example seen in figure \ref{FIG8}, where data in the three lattices are
compared for the trefoil knot.

We do however see the following qualitative features in our results:
(1) Compressibilities are not monotonic with increasing pressure, but
may rise and fall instead, and may even exhibit two local maxima.  That is,
with increasing pressure the lattice knot may become relatively
more compressible in the sense that fractional change in volume may
increase with pressure in certain pressure ranges.  (2) Generally, in
tables \ref{CompKnot-Box} and \ref{CompKnot-Box-CH}, \ref{CompKnot-Box-FCC}
and \ref{CompKnot-Ve-FCC}, and \ref{CompKnot-Box-BCC}
and \ref{CompKnot-Ve-BCC}, there is a weak pattern of increasing 
compressibility at zero pressure down the tables.  (3) We found that FCC
lattice knots are likely to have larger compressibility with the
volume $V_b$, as seen in figures \ref{ScatterBetaVbSCFCC}
and \ref{ScatterBetaVbFCCBCC}, although there are many exceptions to
this.  This is not the case if the volume $V_e$ is used, as one
may for example see in figure \ref{ScatterBetaVeSCFCC}.

We have also computed several other properties of the lattices knots,
such as for example the minimum values of $V_b$ and $V_e$ in each case.
In addition, the maximum amount of useful work that can be extracted
from a knot if it undergoes a reversible isothermic expansion. This sets
an upper bound on the amount of useful work that van be extracted by
the expansion of a compressed lattice knot. Our results were listed
and discussed (see for example figure \ref{ScatterBetaWVeSC}).

%%%%%%%%%%%%%%%%%%%%%%%%%%%%%%%%%%%%%%%%%%%%%%%%%%%%%%%%%%%%%%%%%%%%%%%%%%%%%%%
%%%%%%%%%%%%%%%%%%%%%%%%%%%%%%%%%%%%%%%%%%%%%%%%%%%%%%%%%%%%%%%%%%%%%%%%%%%%%%%
\begin{table}[h!]
\begin{center}
 \begin{tabular}{||c||c||c|c||c|c||r||r||}
 \hline
 Knot & $\beta(0)$& $p_m$  & $\max\beta(p)$ & loc $p_m$ & loc $\max\beta(p)$ & $\C{W}_K$ & Min $V$ \\
  \hline  
$3_1^+ \# 3_1^+$ &\fns $1.27231$ &\fns $0.33117$ &\fns $2.24876$ &\fns $-$ &\fns $-$  &\fns $7.15149$ &\fns $24$ \\
$3_1^+ \# 3_1^-$ &\fns $1.51949$ &\fns $0.20408$ &\fns $1.25617$ &\fns $-$ &\fns $-$  &\fns $8.42508$ &\fns $24$ \\
$3_1^+ \# 4_1$   &\fns $1.76227$ &\fns $0.00000$ &\fns $1.76227$ &\fns $-$ &\fns $-$  &\fns $4.79472$ &\fns $36$ \\
$4_1 \# 4_1$     &\fns $2.30417$ &\fns $0.03292$ &\fns $2.57205$ &\fns $-$ &\fns $-$  &\fns $6.77072$ &\fns $45$ \\
$3_1^+ \# 5_1^+$ &\fns $0.46721$ &\fns $0.19779$ &\fns $2.91011$ &\fns $-$ &\fns $-$  &\fns $6.03715$ &\fns $40$ \\
$3_1^+ \# 5_1^-$ &\fns $1.50415$ &\fns $0.10167$ &\fns $1.73399$ &\fns $-$ &\fns $-$  &\fns $7.07741$ &\fns $40$ \\
$3_1^+ \# 5_2^+$ &\fns $2.22590$ &\fns $0.00000$ &\fns $2.22590$ &\fns $-$ &\fns $-$  &\fns $9.99405$ &\fns $40$ \\
$3_1^+ \# 5_2^-$ &\fns $0.12513$ &\fns $0.24702$ &\fns $0.66874$ &\fns $-$ &\fns $-$  &\fns $2.90872$ &\fns $48$ \\
 \hline
$3_1^+\# 3_1^+$ &\fns $0.04653$ &\fns $0.89401$ &\fns $0.71536$ &\fns $-$ &\fns $-$              &\fns $6.45834$ &\fns $53/4$ \\
$3_1^+\# 3_1^-$ &\fns $0.05839$ &\fns $1.28041$ &\fns $0.13051$ &\fns $-$ &\fns $-$              &\fns $8.42508$ &\fns $97/6$ \\
$3_1^+ \# 4_1$  &\fns $0.05482$ &\fns $0.96456$ &\fns $0.07746$ &\fns $-$ &\fns $-$              &\fns $8.22871$ &\fns $259/12$ \\
$4_1 \# 4_1$    &\fns $0.06322$ &\fns $1.42818$ &\fns $0.09618$ &\fns $-$ &\fns $-$              &\fns $8.85016$ &\fns $51/2$ \\
$3_1^+ \# 5_1^+$&\fns $0.04639$ &\fns $1.28800$ &\fns $0.11847$ &\fns $-$ &\fns $-$              &\fns $8.33974$ &\fns $70/3$ \\
$3_1^+ \# 5_1^-$&\fns $0.04577$ &\fns $1.40675$ &\fns $0.10193$ &\fns $-$ &\fns $-$ &\fns $9.38000$ &\fns $91/4$ \\
$3_1^+ \# 5_2^+$&\fns $0.07706$ &\fns $1.31128$ &\fns $0.16266$ &\fns $-$ &\fns $-$ &\fns $11.24682$ &\fns $70/3$ \\
$3_1^+ \# 5_2^-$&\fns $0.03321$ &\fns $0.95426$ &\fns $0.05034$ &\fns $-$ &\fns $-$ &\fns $4.00733$ &\fns $101/4$ \\
  \hline
 \end{tabular}
\end{center}
 \caption{Compressibility of Compound SC Lattice Knots with volumes $V_b$ (top data) and $V_e$ (bottom data).}
  \label{CompKnot-Compound-SC} %ZXZ[CompKnot-Compound-SC]
\end{table}
%%%%%%%%%%%%%%%%%%%%%%%%%%%%%%%%%%%%%%%%%%%%%%%%%%%%%%%%%%%%%%%%%%%%%%%%%%%%%%%
%%%%%%%%%%%%%%%%%%%%%%%%%%%%%%%%%%%%%%%%%%%%%%%%%%%%%%%%%%%%%%%%%%%%%%%%%%%%%%%

As a final set of calculations, we considered minimal length lattice
knots with compound knot type in the SC lattices.  Our results are
listed in table \ref{CompKnot-Compound-SC}.  Note that in this short list
for both choices of the volume ($V_b$ or $V_e$) that $3_1^+\# 5_2^+$
is the most compressible and has the largest value for $\C{W}_K$.  Similarly,
$3_1^+ \# 5_2^-$ is the least compressible with the lowest value 
of $\C{W}_K$.

\vspace{1cm}

\noindent{\bf Acknowledgements:} The authors acknowledge support 
in the form of NSERC grants from the Government of Canada.

\vspace*{0.5cm}

\noindent{\bf{Bibliography}}

\vspace*{0.3cm}

\begin{center}

\end{center}

\end{document}